\documentclass[11pt]{article}
\pdfoutput=1
\usepackage{jheppub}
\usepackage{amsfonts}
\usepackage{amsmath}
\allowdisplaybreaks[4]         
\usepackage{amssymb}   
\usepackage{euscript}       
\usepackage{xcolor}          
\usepackage{tensor}     
\usepackage{subcaption}
\usepackage[normalem]{ulem} 

\usepackage{physics}
\usepackage{float}
\numberwithin{equation}{section}
\usepackage{lscape}
\usepackage{subcaption}
\usepackage{tabularx}
\usepackage{bbm}
\usepackage{slashed}
\usepackage{bbold}
\usepackage{comment}
\usepackage[makeroom]{cancel}

\usepackage{tikz}
\usepackage{graphicx}
\usetikzlibrary{calc,fadings,decorations.pathreplacing}
\usepackage{verbatim}

\usepackage{url}
\tikzset{%
  >=latex, % option for nice arrows
  inner sep=0pt,%
  outer sep=2pt,%
  mark coordinate/.style={inner sep=0pt,outer sep=0pt,minimum size=3pt,
    fill=black,circle}%
}

\newcommand{\ba}{\begin{eqnarray}}
	\usepackage{braket}
	\newcommand{\ea}{\end{eqnarray}}

\newcommand{\beq}{\begin{equation}}
\newcommand{\beqa}{\begin{eqnarray}}
\newcommand{\beqar}{\begin{eqnarray*}}

\newcommand{\be}{\begin{equation}}
\newcommand{\D}{\mathcal{D}}

\newcommand{\ok}[1]{\textcolor{magenta}{\textcolor{magenta}{(MD:okay)}}}

\usepackage{comment}

\title{\boldmath  Microstates of AdS$_5$ black holes with hypermultiplets

}

\author[a,b]{Marina David,}
\author[a,b]{Annelien Vekemans}

\affiliation[a]{Instituut voor Theoretische Fysica, KU Leuven,
	Celestijnenlaan 200D, B-3001 Leuven, Belgium \vspace{0.1cm}}
 \affiliation[b]{Leuven Gravity Institute, KU Leuven,
Celestijnenlaan 200D, B-3001 Leuven, Belgium \vspace{0.1cm}}

\emailAdd{marina.david@kuleuven.be, annelien.vekemans@kuleuven.be}

\date{\today}
\abstract{We construct supersymmetric rotating AdS$_5$ black holes in 5d $\mathcal{N}=2$ gauged supergravity coupled to two vector multiplets and a universal hypermultiplet, and verify their microscopic counting from the superconformal index of the dual 4d class $\mathcal{S}$ $\mathcal{N}=1$ SCFTs. From the CFT, we perform the Legendre transform of the index to the microcanonical ensemble. The theories are parametrized by a rational number $z$ which enters into the extremization equations making them more challenging to solve. We present a method to address these difficulties and highlight the subtleties involved. From the gravity perspective, we identify a charged, rotating black hole whose Bekenstein-Hawking entropy matches the prediction from the index for $z=\pm1$. Beyond this value, where hypermultiplet scalars are nonzero, we construct the near-horizon extremal geometry perturbatively around $z= 1$ and verify that the entropy is consistent with the CFT prediction. We discuss the thermodynamics and verify the near-horizon versions of the first law of thermodynamics and the supersymmetric condition. In this setting, our analysis characterizes the first construction of a rotating black hole geometry in a 5d $\mathcal{N}=2$ supergravity theory that contains hypermultiplets.

}

\begin{document} 
	\maketitle
	\flushbottom
%%%%%%%%%%%%%%%%%%%%%%%%%%%%%%

\section{Introduction}

The study of black hole thermodynamics and its microscopic origin has been a challenging endeavour in our understanding of quantum gravity, with the successes of \cite{Strominger:1996sh} providing a first step in this direction. In recent years the focus has shifted to asymptotically Anti-de Sitter (AdS) black holes driven by the AdS/CFT correspondence. This has fueled advances in the study of superconformal field theories (SCFTs) and their holographic dual, providing deeper insight on the microscopic nature of asymptotically AdS black holes \cite{Benini:2016hjo, Benini:2016rke,Choi:2018hmj, Cabo-Bizet:2018ehj, Benini:2018ywd, ArabiArdehali:2019tdm,Honda:2019cio,Cabo-Bizet:2019osg,Kim:2019yrz,Cabo-Bizet:2019eaf,Amariti:2019mgp,GonzalezLezcano2019, Lanir:2019abx,Goldstein:2019gpz,ArabiArdehali:2019orz, Murthy:2020rbd,PhysRevD.105.L021903,Agarwal:2020zwm, Benini:2020gjh,Cabo-Bizet:2020nkr, Cabo-Bizet:2021plf,Cassani:2021fyv,Jejjala:2021hlt,Jejjala:2022lrm,Aharony:2021zkr, Cabo-Bizet:2021jar,Goldstein:2020yvj, Choi:2021rxi,Choi:2023tiq,Cassani:2024tvk}. This extensive body of work has yielded significant progress and provided a novel approach for explorations in quantum gravity. Yet, certain theories with holographic duals remain to be fully understood. This stems from challenges from both the gravitational side, i.e, constructing black hole solutions, as well as from computing partition functions of CFTs.

In this work, we focus on 4d class $\mathcal{S}$ $\mathcal{N}=1$ SCFTs realized from reducing 6d $\mathcal{N}=(2,0)$ SCFTs on a topologically twisted hyperbolic Riemann surface $\Sigma_{\mathfrak{g}}$ with genus $\mathfrak{g}$ \cite{Maldacena:2000mw,Benini:2009mz,Bah:2011vv,Bah:2012dg}. The construction of such CFTs provides a framework for exploring so-called non-Lagrangian theories which are otherwise difficult to study using other methods.

These 4d $\mathcal{N}=1$ SCFTs, which we study on a $S^3\times S^1$ background, describe the worldvolume of a stack of $N$ M5-branes wrapped on $\Sigma_{\mathfrak{g}}$ in such a way that it preserves four or eight supercharges. The topological twist is effectively controlled by a rational number $z$. An important object of interest is the superconformal index, which in the large $N$ limit can be identified with the leading dual gravitational saddles. The Cardy-like limit, where the fugacities are small, picks out the so-called black saddle, and the Legendre transform of the index corresponds to the entropy of the holographically dual asymptotically AdS$_5$ black holes \cite{Choi:2018hmj,Honda:2019cio,ArabiArdehali:2019tdm,Kim:2019yrz,Amariti:2019mgp,Cabo-Bizet:2019osg,Gadde:2020bov,GonzalezLezcano:2020yeb,Goldstein:2020yvj,Amariti:2020jyx,Amariti:2021ubd,Cassani:2021fyv,ArabiArdehali:2021nsx}. At leading order in large $N$, the index takes the simple form
\begin{equation}\label{eq:CFTindexintro}
    \log I = \frac{k_{IJK}\hat{\varphi}^I \hat{\varphi}^J \hat{\varphi}^K}{48\hat{\omega}_1 \hat{\omega}_2},
\end{equation}
where $k_{IJK}$ and $k_{I}$ are the cubic and linear 't Hooft anomalies, and $\hat{\omega}_{1,2}, \hat{\varphi}^{I}$ are the fugacities associated to the angular momenta $J_{1,2}$ and electric charges $Q_I$. The practical execution of the Legendre transform often relies heavily on the form of the index. One usually manipulates the extremization equations to reduce them to a single polynomial equation whose solution directly yields the entropy. This is much more manageable than solving a set of equations which are nonlinear in the charges and fugacities, allowing for a fully analytic expression for the entropy.

In many previously known examples for four-dimensional CFTs, e.g., \cite{Benini:2018ywd,Cabo-Bizet:2018ehj, Choi:2018hmj, Amariti:2019mgp, Cassani:2019mms, Cassani:2024tvk} including in certain studies beyond the leading order in $N$ and in the Cardy-like expansion \cite{GonzalezLezcano:2019nca, Lanir:2019abx}, the resulting extremization equation is of third order. However, the anomalies in our theories of interest introduce more technical difficulty and require a more involved approach to analyzing the extremization equations. Only for $z=\pm1$ does the cubic relation hold, while for arbitrary rational values of $z$, we obtain a sixth order polynomial equation. As could be expected, the entropy then takes on a much more nontrivial form and we explain in detail how to obtain it analytically.

From the gravitational dual, the reduction of 7d U(1)$\times$U(1) gauged supergravity on $\Sigma_{\mathfrak{g}}$ yields a 5d $\mathcal{N}=2$ supergravity theory coupled to two vector multiplets and a universal hypermultiplet as presented in \cite{Szepietowski:2012tb}. The AdS vacua dual to the CFT fixed points were found in \cite{Bah:2011vv,Bah:2012dg}, which extended the previously known subset of vacua by Maldacena-N\'{u}\~{n}ez (MN) \cite{Maldacena:2000mw}. The supersymmetric index of the 4d $\mathcal{N}=1$ SCFTs indicates the existence of a class of black hole solutions in this supergravity theory with electric charges $Q_R$ and $Q_F$, although their construction remains an open problem. In this work we aim to address this gap, extending the previous work of \cite{Bobev:2022ocx} where the consistent truncation to minimal gauged supergravity allowed the black hole solution of \cite{Chong:2005hr} to be uplifted to the 7d supergravity for arbitrary values of $z$. What remains a challenge is to construct solutions that support both R-charge and flavour charge.

The presence of the hypermultiplet significantly complicates the task of constructing rotating and electrically charged black holes. In general, it is not always feasible to find analytic solutions due to the nonlinear nature of the equations of motion. In this work, we restrict to solutions with two equal angular momenta, allowing us to exploit SU(2) symmetry as then the angular dependence in the equations of motion becomes trivial, simplifying our ansatz. Moreover, we take a two-step approach, first studying the case of $z=\pm1$ and then extending to other values of $z$.

For $z=\pm1$, the vacua exhibit enhanced supersymmetry and reduce to the $\mathcal{N}=4$ MN geometries \cite{Maldacena:2000mw,Bah:2011vv,Bah:2012dg}. Moreover, the scalars in the hypermultiplet vanish and the theory reduces to a subsector of the STU model, a gauged supergravity coupled to two vector multiplets with gauge group U(1)$^3$. This connection allows us to identify 
a non-extremal non-supersymmetric black hole solution with nonzero R-charge and flavour charge and equal angular momenta, guided by \cite{Cvetic:2004ny}\footnote{Here we only considered the black hole for $z=+1$, but our results can easily be extended to include $z=-1$.}. In the appropriate limits to supersymmetry and extremality, we find that the on-shell action is in agreement with the CFT index.

Guided by the solution we identified for $z=\pm1$ and the solution to the minimal gauged supergravity theory presented in \cite{Bobev:2022ocx}, we extend the analysis beyond these two cases to construct solutions outside $z=\pm 1$ that carry both R- and flavour charges. The equations of motion depend nontrivially on $z$ and are challenging to solve due to the presence of the hypermultiplet scalars. To mitigate this, we focus on the near-horizon extremal geometry (NHEG) which enjoys an SL$(2,\mathbb{R})$ symmetry and all functions in the ansatz are constant. While this is computationally more straightforward, the on-shell action cannot be computed without access to the global solution making it difficult to compare to the index. Instead, we must compare the Bekenstein-Hawking entropy with the one obtained from the Legendre transform of the index.

We introduce an additional simplification and consider a perturbative expansion around $z=1$ such that the equations of motion are now all algebraic and linear at first order in the expansion parameter $\epsilon$. Upon constructing the NHEG at first order, we find a relatively simple way to express the solution. We compute the angular momenta and electric charges from the NHEG using Komar integrals while the fugacities can be read off from the metric and gauge potential. We can then verify a near-horizon version of the first law of thermodynamics and linear supersymmetry constraint. The Bekenstein-Hawking entropy can be written in terms of the charges and this allows us to swiftly confirm the microscopic counting from the CFT. We emphasize that, in this perturbative regime we remain outside the special limit where the hyperscalar vanishes. Therefore, an agreement on the entropy from both sides of the duality is a highly nontrivial check.

Several remarks are worth noting. Our NHEG is a strong indication that a black hole solution exists in these 5d dual supergravity theories with both R-charge and flavour charge away from $z=\pm 1$. Moreover, the solution has a nontrivial hypermultiplet scalar. While some solutions in supergravity theories with nontrivial hypermultiplets exist, e.g. \cite{Gutperle:2001vw,Hristov:2010eu,Halmagyi:2013sla,Erbin:2014hsa,Chimento:2015rra}, we are not aware of any in the context of 5d rotating and electrically charged black hole solutions.

This paper is organized as follows. In Section~\ref{sec:CFTindex}, we present the index of the 4d class $\mathcal{S}$ $\mathcal{N}=1$ CFTs and perform the Legendre transform to the microcanonical ensemble. We comment on the subtleties and certain properties of the index that allow us to solve the extremization equations analytically. Some additional details are included in Appendix~\ref{app:sexticeq}. In Section~\ref{sec:gravitytheory} we provide the consistent truncation to 5d $\mathcal{N}=2$ gauged supergravity coupled to two vector multiplets and a universal hypermultiplet. Additional details on the truncation can be found in Appendix~\ref{app: consistent truncation to 7d} and Appendix~\ref{app:multipletconventions}. In Section~\ref{sec:z=1bhsolutions}, we focus on the black hole solution for the special case of $z=\pm1$. We compute the thermodynamics and the on-shell action and match with the CFT index. Appendix~\ref{app:STU} contains additional details on the relation to the STU model. In Section~\ref{sec:newbhsolutions} we construct the NHEG perturbatively around $z=1$ and study the thermodynamics from the horizon. We explicitly show up to first order in the expansion that the entropy matches the microscopic prediction provided by the CFT dual. We end with final comments and open questions in Section~\ref{section:conclusions}.

%%%%%%%%%%%%%%%%%%%%%%%%%%%%%%
\section{Four-dimensional class \texorpdfstring{$\mathcal{S}$}{} \texorpdfstring{$\mathcal{N}=1$}{} SCFTs from wrapped M5-branes}\label{sec:CFTindex}
%%%%%%%%%%%%%%%%%%%%%%%%%%%%%%

The 6d $\mathcal{N}=(2,0)$ conformal field theory describes the worldvolume dynamics of a stack of $N$ coincident M5-branes. Although this theory is non-Lagrangian and therefore challenging to study using conventional field theory tools, we can still gain insight from symmetry-based arguments. This includes its anomalies and the $N^3$ scaling of its degrees of freedom.

A common strategy for studying such poorly understood theories is to wrap the branes on a closed complex curve $\mathcal{C}_\mathfrak{g}$ with genus $\mathfrak{g}$ and consider the resulting 4d infrared effective theory. More precisely, the 6d $\mathcal{N}=(2,0)$ theory is wound around a topologically twisted $\mathcal{C}_\mathfrak{g}$, and reduced to the 4d class $\mathcal{S}$ SCFTs \cite{Gaiotto:2009gz,Benini:2009mz, Bah:2011vv,Bah:2012dg}. A constant $\kappa = -1, 0, 1$ determines whether the Riemann surface is hyperbolic, toroidal or spherical, respectively\footnote{We will not consider the case of $\kappa=0$. 
}. This topological twist is characterized by a rational number $z$ and ensures that the CFT preserves at least four supercharges \cite{Bah:2011vv, Bah:2012dg}
\footnote{For $\kappa=+1$, $z$ is restricted to $|z|>1$.}

\begin{equation} \label{eq:zdef}
    z = \frac{\ell}{\mathfrak{g}-1}, \qquad \ell\in \mathbb{Z}.
\end{equation}
The numbers $(N,\mathfrak{g},z)$ therefore parametrize an infinite family of 4d $\mathcal{N}=1$ non-Lagrangian SCFTs, where the special case $z=\pm1$ exhibits enhanced supersymmetry to $\mathcal{N}=2$ \cite{Gaiotto:2009we}. These special points of $z=\pm1$ are dual to the $\mathcal{N}=4$ MN vacua \cite{Maldacena:2000mw}. Another interesting case is $z=0$, first studied in \cite{Benini:2009mz} and is dual to the $\mathcal{N}=2$ MN vacua. These holographic dualities were extended to the entire family of SCFTs parametrized by $(N,\mathfrak{g},z)$ in \cite{Bah:2011vv,Bah:2012dg}.

%%%%%%%%%%%%%%%%%%%%%%%%%%%%%%
\subsection{Superconformal index}
%%%%%%%%%%%%%%%%%%%%%%%%%%%%%%

The superconformal index of the 4d class $\mathcal{S}$ $\mathcal{N}=1$ CFTs on $S^1 \times S^3$ is defined as
\begin{equation}\label{eq:generalQFTindex}
    \mathcal{I}_{\text{QFT}}=\text{Tr}\,e^{\pi\,i(1+n_0)F}e^{-\beta\{\mathcal{Q},\bar{\mathcal{Q}}\}+\hat{\omega}_1 \hat{J}_1 + \hat{\omega}_2 \hat{J}_2 + \hat{\varphi}^I \hat{Q}_I},
\end{equation}
where $\hat{J}_{1,2}$ are the angular momenta, $\hat{Q}_I$ are the electric charges, consisting of both superconformal R-charges and flavour charges. The integer $n_0$ can in principle be any value, but will be set to $\pm1$, as dictated by the dual gravity theory.  The supercharge $\mathcal{Q}$ satisfies
\begin{equation}
    [\hat{J}_1,\mathcal{Q}] = [\hat{J}_2,\mathcal{Q}] = \frac{1}{2}\mathcal{Q}, \qquad [\hat{Q}_I,\mathcal{Q}] = -r_I \mathcal{Q},
\end{equation}
with $r_I=1$ for R-symmetries and $r_I=0$ for flavour symmetries. The fugacities $\hat{\omega}_1,\hat{\omega}_2$ and $\hat{\varphi}^I$ are associated to the angular momenta and electric charges and satisfy a linear constraint
\begin{equation}\label{eq:generallinearconstraint}
    \hat{\omega}_1 + \hat{\omega}_2 - r_I \hat{\varphi}^I = 2\pi i n_0.
\end{equation}
For holographic SCFTs the index is particularly interesting in the large $N$ limit, as it describes the structure of the dual gravitational saddles \cite{Benini:2018ywd,GonzalezLezcano:2019nca,Lanir:2019abx,Cabo-Bizet:2019eaf,Cabo-Bizet:2020nkr,Benini:2020gjh,Copetti:2020dil,Cabo-Bizet:2020ewf,Choi:2021rxi,Aharony:2021zkr,Choi:2023tiq}. The contribution corresponding to supersymmetric AdS$_5$ black holes can be isolated by taking the Cardy-like limit of small $\hat{\omega}_i$. When only the U(1) R-charge is turned on, the superconformal index
\begin{equation}\label{eq:Rsymmetryindex}
    \log \mathcal{I}_{\text{QFT}} = \frac{(\hat{\varphi}^R)^3}{48 \hat{\omega}_1 \hat{\omega}_2}k_{RRR} - \frac{\hat{\varphi}^R\left(\hat{\omega}_1^2 + \hat{\omega}_2^2 - 4\pi^2\right)}{48 \hat{\omega}_1 \hat{\omega}_2}k_R + \ldots,
\end{equation}
is given in terms of the linear and cubic 't Hooft anomalies of the R-charge \cite{Cassani:2021fyv}. As dictated by the holographic duality, the index \eqref{eq:Rsymmetryindex} counts the states of a supersymmetric black saddle which has been successfully verified in \cite{Benini:2018ywd,Cabo-Bizet:2018ehj, Choi:2018hmj} to leading order, and to first order in $N$ in \cite{Bobev:2022bjm,Cassani:2022lrk}. This index, however, is not our main target as in addition to the U(1)$_R$ R-symmetry, the 4d class $\mathcal{S}$ CFTs have a U(1)$_F$ flavour symmetry with an associated fugacity $\hat{\varphi}^F$. It was shown in \cite{Cassani:2024tvk} that the index takes on the following form if multiple Abelian symmetries are present, including both R- and flavour symmetries\footnote{Note that the charge fugacities in \cite{Cassani:2024tvk} are normalized in such a way that $(\hat{\varphi}^I)^{\text{there}} = \frac{1}{2}(\hat{\varphi}^I)^{\text{here}}$}
\begin{equation}\label{eq:generalSCFTindex}
    \log \mathcal{I}_{QFT} = \frac{k_{IJK}\hat{\varphi}^I \hat{\varphi}^J \hat{\varphi}^K}{48\hat{\omega}_1 \hat{\omega}_2}-k_I \hat{\varphi}^I \frac{\hat{\omega}_1^2 + \hat{\omega}_2^2 - 4\pi^2}{48\hat{\omega}_1 \hat{\omega}_2} + \ldots.
\end{equation}
We are ultimately interested in the case with one R-symmetry U(1)$_R$ and one flavour symmetry U(1)$_F$, such that
\begin{equation}\label{eq:SCFTindexsubleading}
    \begin{split}
        \log \mathcal{I}_{QFT} =\, &\frac{k_{RRR}(\hat{\varphi}^R)^3 + 3k_{RRF}(\hat{\varphi}^R)^2 \hat{\varphi}^F + 3k_{RFF}\hat{\varphi}^R (\hat{\varphi}^F)^2 + k_{FFF}(\hat{\varphi}^F)^3}{48\hat{\omega}_1 \hat{\omega}_2}\\
        &\qquad + (k_R \hat{\varphi}^R + k_F \hat{\varphi}^F) \frac{\hat{\omega}_1^2 + \hat{\omega}_2^2 - 4\pi^2}{24\hat{\omega}_1 \hat{\omega}_2} + \ldots.
    \end{split}
\end{equation}
The 't Hooft anomalies
\begin{equation}
    k_{IJK} = \text{Tr}(\hat{Q}_I \hat{Q}_J \hat{Q}_K),\qquad k_I = \text{Tr}\hat{Q}_I,
\end{equation}
can be computed by dimensional reduction of the 6d anomaly polynomial \cite{Gaiotto:2009gz,Benini:2009mz,Alday:2009qq,Bah:2011vv,Bah:2012dg,Bobev:2022ocx}. For the class $\mathcal{S}$ SCFTs of type $A_{N-1}$ they are given by
\begin{align}\label{eq:anomaliesBBBW}
    \begin{aligned}
        k_{RRR} &= \frac{2(\mathfrak{g}-1)}{27z^2}\left[9z^2-1- \kappa(3z^2+1)^{\frac{3}{2}}\right]N^3+\mathcal{O}(N)\,, & k_{RRF} & = \frac{(\mathfrak{g}-1)}{9}\, z\, N\,, \\
        k_{RFF} &= \frac{(\mathfrak{g}-1)}{3}\kappa\sqrt{3z^2+1} N^3 +\mathcal{O}(N)\,, & k_{FFF} & = (\mathfrak{g}-1)\, z\, N^3+\mathcal{O}(N)\,, \\
        k_{R}  & = \frac{\mathfrak{g}-1}{3}\left[4+ \kappa\sqrt{3z^2+1}\right]N\,, & k_F & = (\mathfrak{g}-1)\, z\, N \,.
    \end{aligned}
\end{align}
Note that, unlike for many other quantum field theories, the linear 't Hooft anomalies are of $\mathcal{O}(N)$ in the large-$N$ expansion rather than $\mathcal{O}(1)$, making them ideally suited for studying precision holography. Only the cubic anomalies $k_{RRR}$, $k_{RFF}$ and $k_{FFF}$ are of $\mathcal{O}(N^3)$, such that the index to leading order in $N$ is
\begin{equation}\label{eq:CFTindexleadingorder}
    \log \mathcal{I}_{QFT} =\, \frac{k_{RRR}(\hat{\varphi}^R)^3 + 3k_{RFF}\hat{\varphi}^R (\hat{\varphi}^F)^2 + k_{FFF}(\hat{\varphi}^F)^3}{48\hat{\omega}_1 \hat{\omega}_2}.
\end{equation}
We aim to construct the dominating gravitational saddle with an on-shell action as predicted by \eqref{eq:CFTindexleadingorder}.
%%%%%%%%%%%%%%%%%%%%%%%%%%%%%%
\subsection{Legendre transform to the microcanonical ensemble}\label{sec:entropytrick}
%%%%%%%%%%%%%%%%%%%%%%%%%%%%%%
Building on the progress for the theory with global symmetry U(1)$_R$, we seek to relate the index \eqref{eq:CFTindexleadingorder} to the on-shell action of a Euclidean saddle, whose Lorentzian counterpart is expected to be an asymptotically AdS$_5$ supersymmetric black hole. It is then natural to compute the microstate counting that allows us to predict the Bekenstein-Hawking entropy. To do so, we perform a Legendre transform of the index to the microcanonical ensemble
\begin{equation} \label{eq:extremization}
    S = \text{ext}_{(\hat{\omega}_i,\hat{\varphi}^I,\Lambda)}\left\{ -I - \hat{\varphi}^R \hat{Q}_R - \hat{\varphi}^F \hat{Q}_F - \hat{\omega}_1 \hat{J}_1 - \hat{\omega}_2 \hat{J}_2 - \Lambda(\hat{\omega}_1 + \hat{\omega}_2 - \hat{\varphi}^R - 2\pi i )\right\},
\end{equation}
where $I \equiv \log \mathcal{I}_{QFT}$ and $\Lambda$ is a Lagrange multiplier for the linear constraint \eqref{eq:generallinearconstraint}. The variation of \eqref{eq:extremization} with respect to the chemical potentials leads to the following extremization equations
\begin{equation}\label{eq:extremizationeq}
    -\frac{\partial I}{\partial \hat{\omega}_1} = \hat{J}_1 + \Lambda, \qquad -\frac{\partial I}{\partial \hat{\omega}_2} = \hat{J}_2 + \Lambda, \qquad -\frac{\partial I}{\partial \hat{\varphi}^R} = \hat{Q}_R - \Lambda, \qquad -\frac{\partial I}{\partial \hat{\varphi}^F} = \hat{Q}_F.
\end{equation}
Equations \eqref{eq:generallinearconstraint} and \eqref{eq:extremizationeq} provide us with five equations for $(\hat{\omega}_1,\hat{\omega}_2,\hat{\varphi}^R,\hat{\varphi}^F,\Lambda)$ that must be solved in terms of $(\hat{J}_1,\hat{J}_2,\hat{Q}_R,\hat{Q}_F)$. Most of these equations are nonlinear with respect to the chemical potentials and finding their solution is technically challenging. Nevertheless, due to the specific form of \eqref{eq:CFTindexleadingorder}, we can identify a relation that simplifies obtaining the solution. This relation is given by
\begin{equation}\label{eq:trick}
    \frac{1}{6}\gamma^{IJK}\frac{\partial I}{\partial \hat{\varphi}^I} \frac{\partial I}{\partial \hat{\varphi}^J} \frac{\partial I}{\partial \hat{\varphi}^K} - \frac{\partial I}{\partial \hat{\omega}_1} \frac{\partial I}{\partial \hat{\omega}_2} = 0,
\end{equation} 
for a set of constants $\gamma^{IJK}$ that are symmetric in $(IJK)$. The identity \eqref{eq:trick} has been satisfied for many known examples of 4d CFTs, e.g., \cite{Benini:2018ywd,Cabo-Bizet:2018ehj, Choi:2018hmj, Amariti:2019mgp, Cassani:2019mms, Cassani:2024tvk}, and it is easy to see why we might expect such a relation to hold, since derivatives of $I$ in \eqref{eq:CFTindexleadingorder} with respect to $\hat{\varphi}^I$ schematically look like $\partial_{\hat{\varphi}} I \sim (\hat{\varphi}/\hat{\omega})^2$ while differentiating with respect to $\hat{\omega}_i$ gives $\partial_{\hat{\omega}} I \sim (\hat{\varphi}/\hat{\omega})^3$. Applying the condition \eqref{eq:trick} to the index \eqref{eq:CFTindexleadingorder}, we observe that it only holds if
\begin{equation}\label{eq:easytrickcondition}
    k_{RRR} k_{FFF}^2= - 4k_{RFF}^3.
\end{equation}
Imposing the values of the anomalies \eqref{eq:anomaliesBBBW}, we find following constraint on $z$
\begin{equation}\label{eq:BDVargumentz}
    9z^2 - 1 + \kappa (3z^2 + 1)^{3/2} = 0.
\end{equation}
Of course, this is generally not true. Only for $z=\pm1, \, \kappa=-1$ is \eqref{eq:BDVargumentz} satisfied\footnote{Note that the solution $z=0,\,\kappa=+1$ is not allowed by the restriction $|z|>1$ for $\kappa=+1$. Neither is $\kappa=0$, $z=\pm1/3$ since for $\kappa=0$ we must have $z\in \mathbb{Z}$.}. It follows that, in our setup, the identity \eqref{eq:trick} can only be used to obtain the entropy for hyperbolic Riemann surfaces with $z=\pm1$. From this point onward, we restrict our analysis to hyperbolic Riemann surfaces by setting $\kappa =-1$. 

Let us consider the particular values $z=\pm1$ that solve \eqref{eq:BDVargumentz}, and show how the Legendre transform of the index becomes manageable when a relation like \eqref{eq:trick} holds. First, we observe through Euler's theorem $I = \hat{\omega}_1 \partial_1 I + \hat{\omega}_2 \partial_2 I + \hat{\varphi}^R \partial_R I + \hat{\varphi}^F \partial_F I$ and the extremization equations \eqref{eq:extremizationeq} that the entropy is given by
\begin{equation}\label{eq:simpleentropy}
    S = 2\pi i \Lambda.
\end{equation}
We deduce the $\gamma^{IJK}$ that solve \eqref{eq:trick} given by
\begin{equation}
    \gamma^{RRR} = \frac{32}{3 k_{RRR}}, \qquad \gamma^{RFF} = \frac{8}{3 k_{RFF}}, \qquad \gamma^{FFF} = \frac{16}{3 k_{FFF}},
\end{equation}
and use \eqref{eq:extremizationeq} to show that $\Lambda$ satisfies
\begin{equation}\label{eq:cubicequation}
    p_0 + p_1 \Lambda + p_2 \Lambda^2 + \Lambda^3 = 0,
\end{equation}
with
\begin{equation}
    \begin{split}
        &p_0 = \frac{1}{4k_{FFF}^3}\left[9 k_{FFF} k_{RFF}^3\hat{J}_1 \hat{J}_2 + 4\left( 2k_{RFF}\hat{Q}_F - k_{FFF}\hat{Q}_R \right)\left( k_{RFF}\hat{Q}_F + k_{FFF}\hat{Q}_R \right)^2 \right],\\
        &p_1 = \frac{3}{4 k_{FFF}^2}\left( 3k_{RFF}^3 (\hat{J}_1 + \hat{J}_2) - 4 k_{RFF}^2 \hat{Q}_F^2 + 4 k_{FFF}^2 \hat{Q}_R^2 \right), \qquad p_2 = -3\hat{Q}_R + \frac{9k_{RFF}^3}{4 k_{FFF}^2}.
        %&p_2 = -3\hat{Q}_R + \frac{9k_{RFF}^3}{4 k_{FFF}^2}.
    \end{split}
\end{equation}
Since we are considering $z=\pm1$ the anomalies take the simple form
\begin{equation}\label{eq:z=1anomalies}
    k_{RRR} = \frac{32}{27}(\mathfrak{g}-1)N^3, \qquad k_{RFF} = -\frac{2}{3}(\mathfrak{g}-1)N^3, \qquad k_{FFF} = \pm(\mathfrak{g}-1)N^3,
\end{equation}
where the $\pm$ sign in $k_{FFF}$ corresponds to the choice of $z=\pm1$. The entropy of the black saddle can be computed by solving \eqref{eq:cubicequation} for $\Lambda$ and substituting the result into \eqref{eq:simpleentropy}. Since the entropy must be a real quantity, we analyze the roots of the cubic equation \eqref{eq:cubicequation}, which being third order, has three solutions. 

A real entropy can be achieved in one of two ways. One may restrict the parameter space of solutions to the locus $\operatorname{Im}(S)=0$ which leads to the well-known nonlinear constraint given by $p_0 = p_1 p_2$ \footnote{Supersymmetric and extremal black holes satisfy such a constraint and admit a well-defined Lorentzian solution.}. Then the equation admits one real root and a set of two purely imaginary roots for $\Lambda$, and we choose the root that ensures to a positive and real entropy. This procedure was introduced in \cite{Benini:2018ywd,Cabo-Bizet:2018ehj,Choi:2018hmj}. However, as was shown in \cite{Agarwal:2020zwm,Cabo-Bizet:2020ewf,Beccaria:2023hip,Cabo-Bizet:2023ejm}, the nonlinear constraint is not strictly necessary. Instead, we must consider the two leading complex conjugated saddle points corresponding to $n_0=\pm 1$ which have the same boundary conditions and result in two equally dominant contributions. Then the entropy of the combined saddles is given by a real quantity times an oscillatory phase\footnote{It can be shown that two complex solutions can be guaranteed as long as the discriminant of \eqref{eq:cubicequation} is positive. This is satisfied for the valid ranges of the charges.}. In many cases and for the examples of this work, the leading $N$ behavior of the entropy from both points of view is the same\footnote{As a counterexample, for a black hole probed by a D3-brane, it was shown that there is no such nonlinear constraint that results in a real entropy \cite{Cabo-Bizet:2023ejm}.}. Therefore, we will not comment on the subtleties of this further and we find, regardless of the approach, the entropy as a function of the charges
\begin{equation}\label{eq:z=1CFTentropy}
    \begin{split}
        S &= 2\pi \sqrt{\frac{3}{4 k_{FFF}^2}\left( 3k_{RFF}^3 (\hat{J}_1 + \hat{J}_2) - 4 k_{RFF}^2 \hat{Q}_F^2 + 4 k_{FFF}^2 \hat{Q}_R^2 \right)},\\
        &= \frac{2\pi}{\sqrt{3}}\sqrt{-2(\mathfrak{g} -1)N^3(\hat{J}_1+\hat{J}_2)-4\hat{Q}_F^2+9\hat{Q}_R^2},
    \end{split}
\end{equation}
where in the second line we have set the anomalies equal to the values in \eqref{eq:z=1anomalies}. In Section~\ref{sec:z=1bhsolutions}, we construct the black hole solution for $z=\pm1$ that reproduces the prediction \eqref{eq:z=1CFTentropy} of the microstate counting. In the remainder of this section, we consider the more general case of $z\neq\pm1$, where the relation \eqref{eq:trick} does not hold.

%%%%%%%%%%%%%%%%%%%%%%%%%%%%%%
\subsection{Full entropy for any value of \texorpdfstring{$z$}{}}
%%%%%%%%%%%%%%%%%%%%%%%%%%%%%%

For values of $z$ other than $\pm1$, the third-order identity \eqref{eq:trick} is not satisfied and $\Lambda$ no longer obeys a cubic equation. However, we propose a different relation, one of sixth-order
\begin{equation}\label{eq:newtrick}
    \begin{split}
        &\left[\gamma_{(0)}^{IJK} \left(\frac{\partial I}{\partial \hat{\varphi}^I}\right) \left(\frac{\partial I}{\partial \hat{\varphi}^J}\right) \left(\frac{\partial I}{\partial \hat{\varphi}^K}\right)\right]^2 + \gamma_{(2)}\left(\frac{\partial I}{\partial \hat{\omega}_1}\right)^2 \left(\frac{\partial I}{\partial \hat{\omega}_2}\right)^2 \\
        &+ \left[\gamma_{(1)}^{IJK} \left(\frac{\partial I}{\partial \hat{\varphi}^I}\right) \left(\frac{\partial I}{\partial \hat{\varphi}^J}\right) \left(\frac{\partial I}{\partial \hat{\varphi}^K}\right)\right] \left(\frac{\partial I}{\partial \hat{\omega}_1}\right) \left(\frac{\partial I}{\partial \hat{\omega}_2}\right) = 0,
    \end{split}
\end{equation}
for some set of symmetric constants $\gamma_{(0)}^{IJK}$, $\gamma_{(1)}^{IJK}$ and $\gamma_{(2)}$ where $I,J,K=R,F$. Inserting the CFT index \eqref{eq:CFTindexleadingorder} into this relation, we determine the appropriate values for \eqref{eq:newtrick} to hold
\begin{equation}\label{eq:newtrickcoefficients}
    \begin{aligned}
        \gamma_{(0)}^{RRR} &= -\frac{16}{9}k_{FFF}, & \gamma_{(0)}^{RRF} &= \frac{16}{3}k_{RFF}, & \gamma_{(0)}^{FFF} &= \frac{16}{9}k_{RRR},\\
        \gamma_{(1)}^{RRR} &= -\frac{32}{9}\left( 2k_{RFF}^3 + k_{RRR} k_{FFF}^2 \right), & \gamma_{(1)}^{RRF} &= \frac{32}{3} k_{RRR} k_{RFF} k_{FFF}, \\
        \gamma_{(1)}^{RFF} &= -\frac{64}{3}k_{RRR} k_{RFF}^2, & \gamma_{(1)}^{FFF} &= - \frac{32}{9}k_{RRR}^2 k_{FFF},\\
        \gamma^{(2)} &= k_{RRR}\left(4 k_{RFF}^3 + k_{RRR} k_{FFF}^2 \right),
    \end{aligned}
\end{equation}
which holds for any value of $z$. This results in a sixth-order equation for $\Lambda$
\begin{equation}\label{eq:sexticeq}
    q_0 + q_1 \Lambda + q_2 \Lambda^2 + q_3 \Lambda^3 + q_4 \Lambda^4 + q_5 \Lambda^5 + \Lambda^6 = 0.
\end{equation}
The constants $q_i$ are given in terms of the charges and the anomalies, and their explicit expressions can be found in Appendix~\ref{app:sexticeq}. Unlike the previous case in Section~\ref{sec:entropytrick}, the equation \eqref{eq:sexticeq} is sixth-order, making it harder to analyze its roots. We will therefore take the pragmatic approach and restrict ourselves via a nonlinear constraint.\footnote{We reiterate that, as in the case of \cite{Cabo-Bizet:2018ehj,Benini:2018ywd,Choi:2018hmj} the leading term in the entropy takes the same form, regardless of whether one considers both equally dominating saddle or one saddle with a nonlinear constraint.}
To facilitate this, we rewrite \eqref{eq:sexticeq} in a suggestive way
\begin{equation}\label{eq:factorization}
    \left(\Lambda^2 + X\right)\left( Y_0 + Y_1\Lambda + Y_2 \Lambda^2 + Y_3\Lambda^3 + \Lambda^4\right)=0.
\end{equation}
By comparing the two forms of the polynomials \eqref{eq:sexticeq} and \eqref{eq:factorization}, we can solve for $X$ and $Y_i$. We choose to take $-i \sqrt{X}$ to be the purely imaginary root, given that the following nonlinear constraint on the charges must hold
\begin{equation}\label{eq:fullnonlinearconstraint}
    2q_0q_5^3+2 q_1 q_5 \left(q_3-q_4 q_5\right)=\left(q_3^2+q_2 q_5^2-\left(q_1+q_3 q_4\right) q_5\right) \left(q_3+\sqrt{q_3^2-4 q_1 q_5}\right).
\end{equation}
This allows us to write the entropy as
\begin{equation}\label{eq:fullentropy}
    S = 2\pi \sqrt{X} = \sqrt{2} \pi  \sqrt{\frac{q_3+\sqrt{q_3^2-4 q_1 q_5}}{q_5}}.
\end{equation}
This is the predicted microstate counting which holds for any value of $z$. As a consistency check, we take the special value of $z=\pm 1$ and we find that the sixth order equation can be written as a square of the cubic equation found in \eqref{eq:CFTindexleadingorder}, and that the nonlinear constraint \eqref{eq:fullnonlinearconstraint} reduces to $p_0 = p_1 p_2$. 

The difficulty of finding a cubic relation like \eqref{eq:trick} is connected to the structure of the dual 5d $\mathcal{N}=2$ U(1)$_R\times$U(1)$_F$ gauged supergravity theory. Whenever the moduli space in the gravity theory is a symmetric coset space, the structure constants satisfy a closure relation, see Section~\ref{sec:z=1bhsolutions}. This relation ensures the existence of the constants $\gamma^{IJK}$ in \eqref{eq:trick}. We note that the reverse is not true, that having a cubic order relation does not imply that a closure relation in the supergravity theory exists, see for example \cite{Amariti:2019mgp}. When hypermultiplets are included, the scalar manifold is no longer symmetric and the closure relation does not hold. Fortunately, in our case one can still find an identity like \eqref{eq:newtrick}. However, it is not clear how the existence of such a relation is connected to properties of the QFT.

We aim to verify the predictions for the on-shell action and entropy we have presented by constructing black hole solutions in the dual gravity theory. In the following section, we introduce the truncation of 11d supergravity to 5d $\mathcal{N}=2$ supergravity, which will be the theory we focus on for the remainder of this work.

%%%%%%%%%%%%%%%%%%%%%%%%%%%%%%
\section{Consistent truncation from 7d to 5d gauged supergravity}\label{sec:gravitytheory}
%%%%%%%%%%%%%%%%%%%%%%%%%%%%%%

The reduction of 11d gauged supergravity on $S^4$ gives rise to a maximal SO(5) gauged supergravity in seven dimensions \cite{Pilch:1984xy,Nastase:1999cb, Nastase:1999kf, Cvetic:1999xp, Cvetic:1999pu}. For the purposes of this work, we shall focus on an additional truncation  where only the metric with line element $ds_7^2$, a 3-form $\mathcal{S}$, two Abelian gauge potentials $\mathcal{A}_{A,B}$ associated with the U(1)$\times$U(1) Cartan subgroup of $SO(5)$, and two scalars $\lambda_{A,B}$ remain \cite{Liu:1999ai}. We refer the reader to Appendix~\ref{app: consistent truncation to 7d}  for the explicit details of the direct truncation from 11d as they are not essential for the main discussion. Upon reduction, the resulting 7d action is given by
\begin{equation}\label{eq:7DSUGRAaction1}
    \begin{split}
        I_{7} & = \frac{1}{16\pi G_7}\int \Bigg[  \star_7 (R_7 - V_7) - 5 \dd{\lambda_+}\wedge \star_7 \dd{\lambda_+} - \dd{\lambda_-}\wedge \star_7 \dd{\lambda_-} - \frac{1}{2} e^{-4\lambda_A}\mathcal{F}_A\wedge \star_7 \mathcal{F}_A\\
        &\qquad \qquad \quad - \frac{1}{2} e^{-4\lambda_B}\mathcal{F}_B\wedge \star_7 \mathcal{F}_B - \frac{1}{2} e^{-4\lambda_+} \mathcal{S}\wedge \star_7 \mathcal{S} + \frac{1}{2m} \, \mathcal{S}\wedge \dd{\mathcal{S}}\\
        &\qquad \qquad \quad - \frac{1}{m} \mathcal{F}_A\wedge \mathcal{F}_B \wedge \mathcal{S} + \frac{1}{4m}\qty(\mathcal{A}_A\wedge \mathcal{F}_A\wedge \mathcal{F}_B\wedge \mathcal{F}_B + \mathcal{A}_B \wedge \mathcal{F}_B \wedge \mathcal{F}_A \wedge \mathcal{F}_A) \Bigg],
    \end{split}
\end{equation}
where $d\mathcal{A}_{A,B}=\mathcal{F}_{A,B}$, the 7d Hodge dual is denoted by $\star_7$ and $m$ is the coupling. We have defined $ \lambda_\pm \equiv \lambda_A \pm \lambda_B$ and $ V_7$ is the potential for the scalar fields
\begin{equation}\label{eq-scalar-potential1}
    V_7 = \frac{m^2}{2} \qty(e^{-8 \lambda _A-8 \lambda _B}-4 e^{-2 \lambda _A-4 \lambda _B}-4 e^{-4 \lambda _A-2 \lambda _B}-8 e^{2 \lambda _A+2 \lambda _B}) \,.
\end{equation}
The Einstein equation and equations of motion for the scalars and antisymmetric forms can be found in Appendix~\ref{app: consistent truncation to 7d}. The theory described in \eqref{eq:7DSUGRAaction1} is interesting in its own right and admits families of rotating and electrically charged asymptotically AdS$_7$ black hole solutions \cite{Chong:2004dy, Chow:2007ts, Chow:2011fh, Wu:2011gp, Bobev:2023bxl}. In the context of AdS/CFT, these solutions give us access to explicit predictions of observables in 6d CFTs, especially when field theoretical tools are lacking.

Our goal is to contribute to these developments by constructing novel supersymmetric black holes solutions which arise from the backreaction of M5-branes wrapped on a Riemann surface $\Sigma_{\mathfrak{g}}$. The reduction yields a family of 5d $\mathcal{N}=2$ gauged supergravity theories \cite{Szepietowski:2012tb, Faedo:2019cvr,Cassani:2020cod,Bobev:2022ocx}, characterized by $z$ satisfying \eqref{eq:zdef}. It generalizes the truncation to minimal 5d $\mathcal{N}=2$ gauged supergravity \cite{Bobev:2022ocx}, which allows for the construction of black holes with two independent electric charges. The ansatz for the reduction of the 7d metric is
\begin{equation}
    \begin{split} \label{eq:generaltruncationmetric}
        &ds_7^2 = \frac{e^{2f_0 + 4 g_0/3}}{2^{2/3}}(\Sigma e^\varphi)^{4/5}ds_5^2 + \frac{L^2}{4}(\Sigma e^\varphi)^{-6/5}ds_{\Sigma_\mathfrak{g}}^2\,,
    \end{split}
\end{equation}
for the scalar fields, we have
\begin{equation} \label{eq:generaltruncationscalars}
    e^{5\lambda_+} = \Sigma^3 e^{-2\varphi}, \qquad \lambda_- = \alpha,
\end{equation}
and the 1-forms and 3-form are given by
\begin{equation}\label{eq:generaltruncationpforms}
    \begin{split}
        &\mathcal{S} = \frac{e^{f_0}L^2}{32\sqrt{3}}\star_7\left((3z F^F- 16 e^{f_0/2+2g_0}F^R)\wedge d\omega\right),\\
        &\mathcal{A}_A = \frac{L(z+1)}{4}\omega - \frac{1}{8\sqrt{3}}(3A^F+ 4 e^{f_0 + 2\lambda_A^{0}} A^R), \\
        &\mathcal{A}_B = \frac{L(z-1)}{4}\omega + \frac{1}{8\sqrt{3}}(3A^F- 4 e^{f_0 + 2\lambda_B^{0}} A^R),
    \end{split}
\end{equation}
where $m$ is related to the radius of AdS$_5$ $L$ by $m=2/L$. Note that we have chosen to write the ansatz in terms of $f_0,g_0$ and $\lambda_{\pm}^{0}$, which characterize the vacua found in \cite{Bah:2011vv,Bah:2012dg}. Explicitly, they take the form\footnote{These are also the scalar values in the limit to minimal gauged supergravity.}
\begin{equation}\label{eq:BDVansatz2}
    \begin{aligned}
        e^{10\lambda_A^0} &= \frac{1+ 7z + 7z^2 +33 z^3 - (1+4 z + 19 z^2) \sqrt{1+ 3z^2}}{4z(1-z)^2}, & 
        e^{-2\lambda_B^0} &= e^{-2\lambda_A^0} \frac{1+z}{2z +\sqrt{1+3z^2}}, \\
        e^{2g_0} &= \frac{1}{8} e^{2(\lambda_A^0+\lambda_B^0)} \left[(1-z)e^{2\lambda_A^0} + (1+z) e^{2\lambda_B^0}\right] & f_0 &=  4(\lambda_A^0+\lambda_B^0), & 
    \end{aligned}
\end{equation}
After substituting the reduction ansatz into the 7d equations of motion \eqref{eq:7deom}, we find the 5d equations of motion that can be derived from the following action
\begin{equation}\label{eq:5dactionnotheta1}
    \begin{split}
        I_5 = \frac{1}{16\pi G_5}\int \bigg[ &(R_5 - 2 V_5) \star_5 1 - \frac{3}{\Sigma^2} d\Sigma \wedge \star_5 d\Sigma 
        - d\alpha \wedge \star_5 d\alpha - 2 d\varphi \wedge \star_5 d\varphi \\
        &-\frac{1}{2} a_{IJ} F^I \wedge \star_5 F^J - c_{IJK} A^I \wedge F^J \wedge F^K \bigg] \quad \text{where } I, J, K \in (R, F).
    \end{split}
\end{equation}
This theory corresponds to a 5d $\mathcal{N}=2$ U(1)$_R\times$U(1)$_F$ gauged supergravity coupled to two vector multiplets and one hypermultiplet with four real hyperscalars, that parametrize a quaternionic Kahler manifold. Three of these scalars are trivial in this reduction, leaving only one, in particular $\varphi$, with a non-trivial profile. We show in detail how our conventions in \eqref{eq:5dactionnotheta1} are related to the more standard way of writing a 5d theory with $n_{V}$ vector multiplets and $n_{H}$ hypermultiplets in Appendix~\ref{app:multipletconventions}.

The action \eqref{eq:5dactionnotheta1} contains two gauge fields $A^{R,F}$ belonging to the gravity and vector multiplets and two scalars $\alpha$ and $\Sigma$ from the two vector multiplets. The symmetric matrix $a_{IJ}$ characterizes the coupling between the kinetic terms
\begin{equation}
    \begin{split}
    a_{RR} &= \tfrac{\gamma}{3}\left(e^{4\lambda_A^0-2 \alpha }+e^{4\lambda_B^0 + 2 \alpha }+16 \Sigma ^6 e^{4 g_0-f_0}\right), \qquad \quad  a_{FF} = \tfrac{3 e^{-2 f_0}\gamma}{16} \left(e^{-2\alpha}+ e^{2 \alpha}+z^2 \Sigma^6\right),
    \\
    a_{RF}&= \tfrac{e^{-2 \alpha -f_0}\gamma}{4} \left(e^{2\lambda_A^0} - e^{4 \alpha + 2\lambda_B^0 }\right) - \gamma z e^{-3f_0/2+2g_0} \Sigma ^6,
    \end{split}
\end{equation}
where $\gamma = 2^{-4/3}\Sigma^{-2}e^{- 4g_0/3}$. The $c_{IJK}$ are symmetric in the indices and can be mapped to the QFT anomalies \cite{Witten:1998qj,Benvenuti:2006xg}  in \eqref{eq:anomaliesBBBW} via
\begin{equation}
    c_{IJK} = \frac{\sqrt{3} G_5}{4 \pi  L^3}k_{IJK}.
\end{equation}
The scalar potential $V_5$ reads
\begin{equation}\label{eq:5dpotential}
    \begin{split}
        V_5 &= \frac{2^{1/3} e^{2f_0 + 4g_0 /3}L^2}{4 \Sigma^4} \left[
        2 e^{4 \varphi} - 8 \left(e^{-\alpha} + e^{\alpha}\right) \Sigma^3 e^{2 \varphi} \right. \\&
        \qquad \qquad \qquad \qquad \qquad \quad \left. + \Sigma^6 \left(8 e^{2 \varphi} + \tfrac{1}{2} e^{4 \varphi - 2 \alpha} 
        \left(e^{4 \alpha} (z-1)^2 + (z+1)^2\right) - 16\right)
        \right].
    \end{split}
\end{equation}
From the action \eqref{eq:5dactionnotheta1} we derive the scalar equations of motion
\begin{equation} \label{eq:EoMscalar}
    \begin{split}
        &-2\frac{\partial V_5}{\partial\alpha}\star_5 1 + 2d\star_5 d\alpha - \frac{1}{2}\frac{\partial a_{IJ}}{\partial\alpha}F^I\wedge\star_5 F^J = 0,\\
        & -2\frac{\partial V_5}{\partial\Sigma}\star_5 1 + \frac{6}{\Sigma^3}d\Sigma\wedge\star_5 d\Sigma + d\left( \frac{6}{\Sigma^2}\star_5 d\Sigma\right) - \frac{1}{2}\frac{\partial a_{IJ}}{\partial\Sigma}F^I\wedge\star_5 F^J = 0,\\
        & -2\frac{\partial V_5}{\partial\varphi}\star_5 1+ 4d\star_5 d\varphi = 0,
    \end{split}
\end{equation}
the Maxwell equations
\begin{equation} \label{eq:EoMsmaxwell}
    \begin{split}
        d\left(a_{RI} \star F^{I}\right) + 3 c_{RJK} F^{J} \wedge F^{K}=0, \quad
        d\left(a_{FI} \star F^{I}\right) + 3 c_{FJK} F^{J} \wedge F^{K}=0,
    \end{split}
\end{equation}
and the Einstein equations
\begin{equation} \label{eq:EoMseinstein}
    \begin{split}
        R_{5,\mu\nu} &= \frac{3}{\Sigma^2}\partial_\mu\Sigma \partial_\nu \Sigma + \partial_\mu\alpha \partial_\nu \alpha +2\partial_\mu\varphi \partial_\nu \varphi\\
        &+ \frac{1}{2}a_{IJ} F^I_{\mu\lambda} F^{J\lambda}_\nu + \frac{1}{2}g_{\mu\nu}\left[ \frac{4}{3}V_5 - \frac{1}{6}a_{IJ}F^I_{\mu\nu} F^{J\mu\nu}\right].
    \end{split}
\end{equation}
With the theory at hand, we are now equipped to analyze and construct black hole solutions offering insights on the structure of \eqref{eq:CFTindexleadingorder} from a holographic point of view. Before proceeding, we highlight a few important points.

The additional truncation of \eqref{eq:5dactionnotheta1} to the subgroup U(1)$_R$, with the three scalar fields set to the constant values pertaining to the vacua found in \cite{Bah:2011vv,Bah:2012dg}
\begin{equation}\label{eq:BBBW_scalars}
    \begin{split}
        e^{2\alpha} &= e^{2\lambda_-^0} = \frac{1+3 z-\sqrt{3 z^2+1}}{-1+3z+\sqrt{3 z^2+1}},
        \qquad \qquad \qquad e^{2\varphi} = \frac{1}{4}e^{-2g_0 - \frac{1}{2}f_0} = \frac{4}{\sqrt{1+3z^2}+2}, \\
        \Sigma^{-3} &= \frac{1}{4}e^{-2g_0 + \frac{3}{4}f_0} = \sqrt{\frac{1+z^2+\sqrt{3 z^2+1}}{2}},
    \end{split}
\end{equation}
reduces the theory to 5d $\mathcal{N}=2$ minimal gauged supergravity. Black hole solutions of this theory have been extensively studied \cite{Gutowski:2004ez, Cvetic:2004ny, Chong:2005hr} with the most general non-supersymmetric black hole solutions characterized by two unequal angular momenta and one R-charge. These black holes can be embedded into the reduction ansatz \eqref{eq:generaltruncationmetric}, \eqref{eq:generaltruncationscalars} and \eqref{eq:generaltruncationpforms}. This was carefully studied and a full holographic analysis was performed in \cite{Bobev:2022ocx}. Unfortunately, this class of black hole solutions will not suffice to study the full structure of the index \eqref{eq:CFTindexleadingorder}. To address this, we refrain from this additional truncation and consider the full gauge group U(1)$_R\times$U(1)$_{F}$.

The MN vacua first identified in \cite{Maldacena:2000mw}, are a subset of the more general vacua identified in \cite{Bah:2011vv,Bah:2012dg} with the former are valid for arbitrary rational values of $z$. Among the MN vacua, the case of $z=\pm1$ is particularly interesting, since they have enhanced supersymmetry to $\mathcal{N}=4$. Moreover, at this value, the hyperscalar $\varphi$ vanishes and the theory is a subsector of the STU model. This allows us to identify black hole solutions with two independent electric charges as studied in \cite{Gutowski:2004yv,Cvetic:2004ny,Cvetic:2005zi,Kunduri:2006ek}. We describe this solution in detail in Section~\ref{sec:z=1bhsolutions}.

Relaxing the value of $z=\pm 1$ outside the regime of enhanced supersymmetry, our second goal is to construct black hole solutions for all rational values of $z$ which are predicted to have an entropy given by \eqref{eq:fullentropy} and asymptote to the AdS$_5$ vacua described in \eqref{eq:BBBW_scalars}. Since the supergravity theory has gauge group U(1)$_R\times$U(1)${}_F$, the most general solution has two independent electric charges $Q_R$ and $Q_F$. One of the difficulties in obtaining these solutions arises due to the nontrivial profile of the hyperscalar. Exploring the construction of these solutions is the main focus of Section~\ref{sec:newbhsolutions}.

%%%%%%%%%%%%%%%%%%%%%%%%%%%%%%
\section{Black holes for \texorpdfstring{$z=\pm1$}{}}\label{sec:z=1bhsolutions}
%%%%%%%%%%%%%%%%%%%%%%%%%%%%%%

%%%%%%%%%%%%%%%%%%%%%%%%%%%%%%
\subsection{The black hole solution} \label{sec:z=1explicitsol}
%%%%%%%%%%%%%%%%%%%%%%%%%%%%%%
For the special value of $z=\pm1$, the theory \eqref{eq:5dactionnotheta1} admits $\mathcal{N}=4$ MN vacua, a special case of those found in \cite{Bah:2011vv,Bah:2012dg}. In this limit
\begin{equation}
    \Sigma^3 = \frac{1}{\sqrt{2}}, \qquad e^{\pm\alpha} = \frac{1}{\sqrt{2}}, \qquad e^{2\varphi} = 1,
\end{equation}
and the two scalars in the vector multiplets $\Sigma$ and $\alpha$ are constrained via
\begin{equation}\label{eq:z=1scalars}
    e^{\pm\alpha} = \Sigma^3.
\end{equation}
In the following, we allow $\alpha$ and $\Sigma$ to flow according to \eqref{eq:z=1scalars} and set $\varphi$ to its vanishing vacuum value, effectively truncating the hypermultiplet. With these conditions, we construct black hole solutions to the equations of motion \eqref{eq:EoMscalar}, \eqref{eq:EoMsmaxwell} and \eqref{eq:EoMseinstein}. For simplicity, we cover only the $z=+1$ case\footnote{Our results are easily generalizable to $z=-1$ with the help of Appendix~\ref{app:STU}.} and restrict to two equal angular momenta, as then the angular dependence of the ansatz becomes trivial. In deriving this solution, we were guided by the known one-charge ($Q_{R} \neq 0, Q_F=0$) black holes from the minimal truncation \cite{Bobev:2022ocx} and a relation to the STU model, which has gauge group U(1)$^3$.
In this case, two of the gauge fields of the STU model become equal and a particular linear combination of the remaining gives the gauge potentials of the R-symmetry and flavour symmetry. We refer the reader to Appendix~\ref{app:STU} for details.  Considering the above, the metric reads
\begin{equation}\label{eq:z=1metric5d}
\text ds^2 = \left(H_1 H_2^2\right)^{1/3}\left[ -\frac{r^2 Y}{f_1}\text dt^2+ \frac{r^4}{Y}\text dr^2+\frac{r^2}{4}\left(\sigma_1^2 + \sigma_2^2\right)+\frac{f_1}{4r^4H_1H_2^2}\left(\sigma_3 -\frac{2f_2}{f_1}\text dt\right)^2
\right]\,,
\end{equation}
which is expressed in terms of the following functions
\begin{align}
\begin{split}
H_i(r) &= 1+\frac{2m \,s_i^2}{r^2}, \\
f_1(r)&= 4 a^2 m^2 \left[2\,s_1 s_2^2 \left(c_1c_2^2-s_1s_2^2\right)-2s_1^2 s_2^2-s_2^4\right]+2 a^2 m\, r^2+H_1H_2^2\,  r^6, \\
f_2(r) &= 2m\,a\left(c_1c_2^2-s_1s_2^2\right)r^2+ 4m^2as_1s_2^2\,, \\
f_3(r)&= 4 m^2 \frac{a^2}{L^2} \left[2\, s_1 s_2^2 \left(c_1c_2^2-s_1s_2^2\right)-2s_1^2 s_2^2-s_2^4\right]+2 a^2 m \left(\frac{r^2}{L^2} +1\right)\,, \\
Y(r) &= f_3(r) + \frac{r^6}{L^2} H_1H_2^2 + r^4 - 2m\, r^2\,,
\end{split}
\end{align}
with $i=1,2$. The metric is written using Maurer-Cartan 1-forms that parametrize SU(2) which are given by
	\begin{align} \label{eq:left1forms}
		\begin{aligned}
			\sigma_1 & =\sin \phi d \theta-\cos \phi \sin \theta d \psi, \quad
			\sigma_2 =\cos \phi d \theta+\sin \phi \sin \theta d \psi, \quad
			\sigma_3 =d \phi+\cos \theta d \psi.
		\end{aligned}
	\end{align}
The Euler angles have a range of $0 \leq \theta \leq \pi, 0 \leq \phi \leq 4\pi, 0 \leq \psi \leq 2\pi$. The solution is characterized by four independent parameters $(m, a, \delta_1, \delta_2 )$ which are associated to the mass, angular momentum and two charges. It is convenient to rewrite $\delta_I$ with hyperbolic functions where $s_I = \sinh \delta_I $ and $c_I = \cosh \delta_I$. Additionally, it may at times be easier to write certain quantities in terms of the outer horizon $r_+$ which is the largest positive root of the function $Y(r)$. However, it is not an independent parameter. The two gauge potentials are given by
\begin{equation}\label{eq:z=1gaugefields}
    \begin{split}
        A^R &=  z_R dt + \frac{2m}{\sqrt{3}r^2 \overline{H}}\left[\left( 2c_2 s_2 H_1 + c_1 s_1 H_2 \right)dt - \frac{a}{2}\left( 2 c_2 s_2 (c_1 - s_1) H_1 + (c_2^2 s_1 - s_2^2 c_1) H_2 \right)\sigma_3\right],\\
        A^F &= z_F dt +\frac{8m}{3\sqrt{3}r^2 \overline{H}}\left[\left( -c_2 s_2 H_1 + c_1 s_1 H_2 \right)dt + \frac{a}{2}\left(c_2 s_2 (c_1 - s_1) H_1 - (c_2^2 s_1 - s_2^2 c_1) H_2 \right)\sigma_3\right],
    \end{split}
\end{equation}
where $\overline{H}\equiv H_1 H_2$. We include pure gauge terms $z_{R,F} dt$ which will be fixed later to ensure regularity of $A^{I\,\mu}A^I_\mu$ at the horizon. The scalars are given by 
\begin{equation} \label{eq:z=1scalars1}
    \Sigma^6 = \frac{H_2}{2H_1}, \quad e^\alpha = \sqrt{\frac{H_2}{2H_1}}, \quad \varphi = 0.
\end{equation}
Below, we review the thermodynamics of the solution and its supersymmetric limit and show with what is now the standard holographic procedure, the equivalence between the on-shell action and the CFT index.

%%%%%%%%%%%%%%%%%%%%%%%%%%%%%%
\subsection{Thermodynamics and holographic match}
%%%%%%%%%%%%%%%%%%%%%%%%%%%%%%
To fully characterize the solution, we first compute its thermodynamic quantities which provide the necessary input for studying the on-shell action and entropy in a holographic setting. This will enable us to verify the CFT prediction for the on-shell action and likewise the entropy as proposed in Section~\ref{sec:CFTindex}. We start with the black hole entropy which is given by the area formula
\begin{align}
    S = \frac{\sqrt{f_1(r_+)}\pi^2}{2G_5}.
\end{align}
The inverse temperature $\beta$ can be determined by Wick rotating the solution at the horizon, and requiring the absence of a conical singularity. The periodicity of the Euclidean time coordinate then determines the inverse temperature, giving
\begin{align}
    T = \beta^{-1} = \frac{Y(r_+)}{4\pi r_+ \sqrt{f_1(r_+)}}.
\end{align}
The angular momentum $J$ can be computed by performing a Komar integral over the $S^3$ at infinity
\begin{align}
    J = -\frac{1}{16\pi G_5} \int_{\partial\mathcal{M}} \star dK,
\end{align}
where $K=-g_{\mu\phi}dx^{\mu}$ is dual to the Killing vector $\xi = -\partial_{\phi}$.
We compute the so-called Maxwell charges $Q_{R,F}$ 
\begin{equation}
    Q_I = -\frac{1}{16\pi G_5} \int_{\partial\mathcal{M}} a_{IJ} \star F^J,
\end{equation}
where the integration is performed over the null surface at infinity and should asymptotically coincide with the other notions of charge,  provided that they decay fast enough \cite{Marolf:2000cb}. The angular velocity can be obtained by identifying the Killing vector $\ell^{\mu}=\partial_t+\Omega \partial_{\phi}$ that becomes null at the horizon
\begin{align}
    \Omega = \frac{2f_2(r_+)}{f_1(r_+)}.
\end{align}
The electrostatic potentials $\Phi^{R,F}$ are defined as
\begin{align}
    \Phi^{I} = \left.\ell^{\mu} A^{I}_{\mu} \right|_{r=r_+} -\left.\ell^{\mu} A^{I}_{\mu} \right|_{r \to \infty}. 
\end{align}
To ensure regularity of $A^{I\,\mu}A^I_\mu$ at the horizon, the pure gauge terms must be fixed to
\begin{equation}
    z_R = -\Phi^R, \qquad z_F = -\Phi^F.
\end{equation}
The energy can be computed via the ADM formula  \cite{Ashtekar:1999jx} or via the first law of black hole thermodynamics
\begin{equation}\label{eq:firstlawBH}
    dE = T dS +\Omega dJ + \Phi^I dQ_I,
\end{equation}
where variations are taken with respect to the black hole parameters $(m,a,\delta_1,\delta_2)$\footnote{Or, alternatively, four of the five parameters $(r_+,m, a,\delta_1,\delta_2)$.}. We now present the results of the above computations. The thermodynamic charges are given by
\begin{equation}\label{eq:z=1charges}
    \begin{aligned}
        E &= \frac{\pi}{4G_5}m\left( 3+ \frac{a^2}{L^2} + 2(s_1^2 + 2s_2^2) \right), & \quad J &=\frac{\pi}{2 G_5}m a \left(c_1 c_2^2-s_1 s_2^2\right),\\
        Q_R &= \frac{\pi}{2 \sqrt{3} G_5}  m \left(c_1 s_1+2 c_2 s_2\right), & \quad Q_F &= \frac{\sqrt{3} \pi}{4 G_5}  m \left(c_1 s_1-c_2 s_2\right),
    \end{aligned}
\end{equation}
the entropy reads
\begin{equation}\label{eq:z=1entropy}
    S = \frac{\pi^2 L}{2G_5}\sqrt{2m(r_+^2 - a^2) - r_+^4},
\end{equation}
and the chemical potentials are
\begin{equation}\label{eq:z=1potentials}
    \begin{split}
        &\beta = \frac{2\pi L^3 \sqrt{2m(r_+^2 - a^2) - r_+^4}}{2 m \left( \left(a^2+2 \left(s_1^2+2 s_2^2\right) r_+^2\right)- L^2 \right)+4 m^2 s_2^2\left(2s_1^2+s_2^2\right)+3 r_+^4+2 r_+^2  L^2},\\
        &\Omega = \frac{4a m\left( 2m s_1 s_2^2 + (c_1 c_2^2 - s_1 s_2^2)r_+^2 \right)}{L^2 \left(2m(r_+^2 - a^2) - r_+^4\right)},\\
        &\Phi^R = \frac{m}{\sqrt{3}}\left( \frac{2c_1 s_1}{r_+^2 + 2ms_1^2}+ \frac{4c_2 s_2}{r_+^2 + 2ms_2^2} -a \Omega\left( \frac{c_2^2 s_1 - c_1 s_2^2}{r_+^2 + 2ms_1^2} - \frac{2c_2 s_2(s_1 - c_1)}{r_+^2 + 2ms_2^2} \right)\right),\\
        &\Phi^F = \frac{4m}{3\sqrt{3}}\left( \frac{2c_1 s_1}{r_+^2 + 2ms_1^2} - \frac{2c_2 s_2}{r_+^2 + 2ms_2^2} -a \Omega\left( \frac{c_2^2 s_1 - c_1 s_2^2}{r_+^2 + 2ms_1^2} + \frac{c_2 s_2(s_1 - c_1)}{r_+^2 + 2ms_2^2} \right)\right).
    \end{split}
\end{equation}
Imposing the charges \eqref{eq:z=1charges}, entropy \eqref{eq:z=1entropy} and chemical potentials \eqref{eq:z=1potentials}, we deduce the on-shell action from the quantum statistical relation
\begin{equation}
    I_{5} = \beta E - S - \beta \Omega J - \beta \Phi^I Q_I,
\end{equation}
which is given by
\begin{equation}\label{eq:z=1OSA}
    \begin{split}
        I_{5}^{} = -&\frac{\pi \beta}{12G_5}\left[ 3m\left(\frac{a^2}{L^2} - 1\right) + 3 \frac{r_+^4}{L^2} + 2m \left(2 \frac{r_+^2}{L^2} - 1\right)\left(s_1^2 + 2 s_2 ^2\right)\right.\\
        &\left.+ \frac{4m^2 s_2^2}{L^2} \left(2 s_1^2 + s_2^2\right) + 2m \left(-\frac{c_1 s_1}{2\sqrt{3}}\left(3\Phi^F + 2\Phi^R\right) + \frac{c_2 s_2}{4\sqrt{3}} \left( 3\Phi^F-4\Phi^R \right) \right) \right].
    \end{split}
\end{equation}
This coincides with the more rigorous approach of computing the on-shell action using a suitable renormalization scheme such as holographic renormalization or background subtraction, as was implemented in \cite{Cassani:2019mms,Cassani:2024tvk}.
We are ultimately interested in the supersymmetric limit of the solution, which must satisfy
\begin{align}\label{eq:z=1susyconditioncharges}
    E = \frac{2}{L} J +\sqrt{3} Q_R,
\end{align}
where only the R-charge appears in \eqref{eq:z=1susyconditioncharges} since flavour symmetries commute with the superalgebra. Substituting the thermodynamic quantities found in \eqref{eq:z=1charges}, we find that this sets a constraint on the parameters of the black hole solution
\begin{align}\label{eq:z=1susyconditionparams}
    e^{\delta_1 + 2\delta_2} = \frac{L}{a}.
\end{align}
Upon imposing the above supersymmetric constraint, the solution and its charges and chemical potentials are complex-valued. It is only when the extremal condition is applied that the reality of the charges is preserved and the Lorentzian black hole solution is well-defined. However, the chemical potentials remain complex.

At extremality, the function $Y(r)$ develops a double root and the temperature of the black hole vanishes. As a consequence of imposing supersymmetry and extremality, the solution can be written in terms of two independent parameters, which we choose to be $\delta_1$ and $\delta_2$. The rotation parameter $a$ is determined by \eqref{eq:z=1susyconditionparams}, while the explicit values of the horizon radius and $m$ read
\begin{equation} \label{eq:extremalhorizon}
    \begin{split}
        r_*^2 &= \frac{L^2 \left(-2 e^{2 \delta _2}+4 e^{2 \left(\delta _1+\delta _2\right)}-2\right)}{\left(e^{2 \delta _2}+1\right) \left(e^{2 \left(\delta _1+\delta_2\right)}-1\right)^2},\qquad
        m_{*} = \frac{4 L^2 e^{2 \delta _1+4 \delta _2}}{\left(e^{4 \delta _2}-1\right) \left(e^{2\left(\delta _1+\delta _2\right)}-1\right)^2}.
    \end{split}
\end{equation}
Here the star denotes that that this is the value when both supersymmetry and extremality are imposed.
In this limit, the chemical potentials take the simple form
\begin{equation}
    \begin{split}
        \Omega^* = \frac{2}{L}, \qquad
        \Phi^R{}^* =\sqrt{3}, \qquad \Phi^F{}^* = 0.
    \end{split}
\end{equation}
To take the supersymmetric and extremal limit in the quantum statistical relation, we define supersymmetric fugacities
\begin{equation} \label{eq:fugacities1}
    \omega \equiv \beta(\Omega - \Omega^*), \qquad \varphi^I \equiv \beta(\Phi^I - \Phi^I{}^*),
\end{equation}
that are manifestly temperature-independent and satisfy the linear constraint\footnote{Note that in principle, the right hand side of \eqref{eq:linearconstraintgrav} should be $\pm 2\pi i$. The different signs here correspond to the freedom to choose a sign when relating the parameters $m$ and $r_+$.}
\begin{equation}\label{eq:linearconstraintgrav} 
    \omega - \frac{\sqrt{3}}{L} \varphi^R = 2\pi i.
\end{equation}
The on-shell action in terms of these fugacitites takes on a very simple, temperature-independent form
\begin{equation}
    I_{5}^{} = \frac{\pi}{3\sqrt{3} G_5}\frac{32 (\varphi^R)^3 - 54 \varphi^R (\varphi^F)^2 + 27 (\varphi^F)^3}{32\omega^2}.
\end{equation}
The quantum statistical is now manifestly temperature-independent and given by
\begin{equation}
    S = -I_5 - \omega J - \varphi^I Q_I.
\end{equation}
Although it is sufficient to verify the on-shell action with the CFT index \eqref{eq:CFTindexleadingorder}, we may also recast the Bekenstein Hawking entropy in terms of the supersymmetric charges
\begin{equation} \label{eq:z=1entropygravity}
    S = \frac{2\pi}{\sqrt{3}}\sqrt{-\frac{3\pi L^3}{4G_5}2 J -4 \left(\frac{L}{\sqrt{3}} Q_F\right)^2 + 9 \left(\frac{L}{\sqrt{3}} Q_R\right)^2}.
\end{equation}
It is now straightforward to compare the above expressions with \eqref{eq:CFTindexleadingorder}, \eqref{eq:extremization} and \eqref{eq:z=1CFTentropy} once we relate the gravity charges $(J, Q_R, Q_F)$ and fugacities $(\omega, \varphi^R, \varphi^F)$ to the QFT counterparts $(\hat{J}_1, \hat{J}_2, \hat{Q}_R, \hat{Q}_F)$ and $(\hat{\omega}_1, \hat{\omega}_2, \hat{\varphi}^R, \hat{\varphi}^F)$ by
\begin{equation} \label{eq:gravityCFTrelation}
    \begin{split}
        \omega = 2\hat{\omega}_1 = 2\hat{\omega}_2, \qquad \varphi^I = \frac{L}{\sqrt{3}} \hat{\varphi}^I, \qquad 
        J = \hat{J}_1 = \hat{J}_2, \qquad Q_I = \frac{\sqrt{3}}{L} \hat{Q}_I.
    \end{split}
\end{equation}
Additionally, the 5d Newton constant $G_5$ can be expressed as
\begin{equation}
    G_5 = \eval{\frac{3\pi L^3}{64 e^{3f_0+2g_0} (\mathfrak{g}-1) N^3}}_{z=1} = \frac{3\pi L^3}{8 (\mathfrak{g}-1) N^3}.
\end{equation}
To conclude, we have identified a black hole solution with equal angular momenta, one R-charge and one flavour charge whose entropy and on-shell action are consistent with the predictions from the dual CFT as given in \eqref{eq:CFTindexleadingorder} and \eqref{eq:z=1CFTentropy}. We note that the construction of this solution was made tractable by setting $z=1$, since then the 5d supergravity theory reduces to a subsector of the STU model.
The scalar fields belonging to the vector multiplet in this model parameterize a symmetric space and the closure relation
\begin{align} \label{eq:closurerelation}
c_{I J K} c_{J^{\prime}(L M} c_{P Q) K^{\prime}} \delta^{J J^{\prime}} \delta^{K K^{\prime}}=\frac{4}{3} \delta_{I(L} c_{M P Q)} ,
\end{align}
is satisfied. The closure relation \eqref{eq:closurerelation} implies the existence of a cubic relation \eqref{eq:trick}, which in turn guarantees that the Bekenstein-Hawking entropy \eqref{eq:z=1entropygravity} matches the entropy predicted from the CFT in \eqref{eq:z=1CFTentropy}.

%%%%%%%%%%%%%%%%%%%%%%%%%%%%%%
\section{Black holes beyond \texorpdfstring{$z=\pm1$}{}}\label{sec:newbhsolutions}
%%%%%%%%%%%%%%%%%%%%%%%%%%%%%%

\subsection{The near-horizon extremal geometry}\label{sec:NHEKsection}

In the following, we construct, at least in part, a more general solution, from which the two black hole geometries found in Section~\ref{sec:z=1bhsolutions} and in minimal gauged supergravity \cite{Bobev:2022ocx} arise as specific limiting cases. Our ultimate goal is to construct solutions of the explicit truncation \eqref{eq:5dactionnotheta1} for non-trivial values of $Q_{R}$ and $Q_{F}$ and for all values of $z$. Once more, we restrict our analysis to the case of equal angular momenta for simplicity, as this leads to a solution with SU(2) symmetry.

Proposing a good ansatz and solving equations of motion is highly involved as the equations of motion depend on the radial coordinate and $z$ in a nontrivial way due to the presence of the hypermultiplet. This makes explicit and analytic solutions challenging to obtain. Despite the difficulty, it is still possible to extract information of this more general black hole without knowing the full geometry. Our approach is to construct the solution of the near-horizon extremal geometry (NHEG), which develops an AdS$_2$ throat due to the enhancement of symmetry to SL$(2,\mathbb{R})$ \cite{Bardeen:1999px}. In this limit, all the functions in the metric, gauge potentials and scalars are constant and the near-horizon region itself becomes an exact solution of Einstein's equations. Taking this into account, we can express the metric ansatz as
\begin{equation} \label{eq:NHEKansatzmetric}
	\begin{split}
		&ds^2 = g_1 \left( -r^2 dt^2 + \frac{dr^2}{r^2} \right) + g_2 \left( \sigma_1^2 + \sigma_2^2 \right) + g_3 \left(\sigma_3 - \widetilde{\omega}\, r dt \right)^2\,,
	\end{split}
\end{equation}
where we have made the SL$(2,\mathbb{R}) \times$ SU(2) symmetry explicit. This enables us to express the metric with an explicit AdS$_2$ factor and use the Maurer-Cartan 1-forms as written in \eqref{eq:left1forms}. The scalars have a nontrivial profile in the full geometry and reach a constant value at the horizon
\begin{align} \label{eq:NHEKansatzscalar}
    \alpha = \widetilde{\alpha}, \quad \Sigma = \widetilde{\Sigma}, \quad \varphi = \widetilde\varphi,
\end{align}
while the ansatz for the gauge fields takes the form
	\begin{align} \label{eq:NHEKansatzpotential}
		A^{R,F} = a_{R,F}\, rdt + b_{R,F}(\sigma_3 - \widetilde{\omega}\, r dt).
	\end{align}
The advantage of the NHEG is that $(g_1,g_2,g_3,\widetilde{\omega},a_{R,F},b_{R,F},\widetilde{\alpha},\widetilde{\Sigma},\widetilde{\varphi})$ are constants and as a result, all the equations of motion are algebraic, greatly simplifying the analysis. However, even with a particular solution of the NHEG, the on-shell action cannot be determined, which is required to compare to the CFT index \eqref{eq:CFTindexleadingorder}. In practice, determining the entropy is sufficient to test the prediction derived from the CFT index, with the ultimate goal of verifying the duality for all rational values of $z$.

Nevertheless, the equations of motion for $z\neq \pm 1$ remain nonlinear and an analytic solution is still technically challenging. To proceed, we consider a linearization around the known solution for $z=1$ given in Section~\ref{sec:z=1bhsolutions}. The approach simplifies the equations of motion which now become both algebraic and linear at each order, providing easy access to the solution and its entropy. We find that this is sufficient in providing (1) evidence for a black hole solution for $z\neq 1$ from the NHEG, (2) the thermodynamic quantities accessible from the NHEG and (3) a confirmation of whether this black hole and its entropy correspond to the dominating saddle and the counting of microstates predicted by the CFT index \eqref{eq:CFTindexleadingorder}.

To proceed, we may either start with the solution of \cite{Cvetic:2004ny,Cvetic:2005zi} and take the smooth supersymmetric and extremal limit, i.e., \eqref{eq:z=1susyconditionparams} and \eqref{eq:extremalhorizon}, or directly with the solution found in \cite{Gutowski:2004yv} provided that the appropriate limit of the STU model is taken as discussed in Appendix~\ref{app:STU}. To approach the NHEG, we implement the following
\begin{align} \label{eq:NHEGlimit}
    r \to r_+ + \frac{r}{\zeta} \quad t \to c_t \frac{t}{\zeta}, \quad \phi \to \phi +c_t \frac{t}{\zeta}, \qquad \zeta \to 0,
\end{align}
where we have rescaled $t$ such that the metric contains a manifest AdS$_2$ factor. Explicitly, we find that the functions in the metric take the form 
	\begin{align} \label{eq:zerosol1}
		g_1=\frac{L^2 (\mu_1\mu_2^2)^{1/3}}{4 \left(\mu _1+2 \mu _2+L^2\right)},
        \quad
        g_2=\frac{1}{4}(\mu_1\mu_2^2)^{1/3}, 
        \quad
        g_3=\frac{(\mu_1^{-1} \mu_2)^{2/3}(4 \mu _1 \left(\mu _2+L^2\right)-\mu _2^2)}{16 L^2},
	\end{align}
and
\begin{align} \label{eq:zerosol2}
    \widetilde{\omega} =-\frac{\left(2 \mu _1+\mu _2\right) L^2}{(4 \mu _1 \left(\mu _2+L^2\right)-\mu _2^2)^{1/2} \left(\mu _1+2 \mu _2+L^2\right)}, \quad c_t = \frac{\mu_2 L \left(\left(4 \mu _1-\mu _2\right) \mu _2+4 \mu _1 L^2\right)^{1/2}}{4 \left(\mu _1+2 \mu _2+L^2\right)}.
\end{align}
The gauge potentials are given by
\begin{equation} \label{eq:zerosol3}
    \begin{aligned}
    a_R &= -\frac{4L^2\left(2 \mu _1+\mu _2\right) c_t}{\sqrt{3} \mu _2 \left(\left(4 \mu _1-\mu _2\right) \mu _2+4 \mu _1 L^2\right)},
    \quad & b_R &=
    \frac{\left(\mu
   _2-4 \mu _1\right) \mu _2}{4 \sqrt{3} \mu _1 L},
   \\
   a_F &=  \frac{8\left(\mu _1-\mu _2\right)(2L^2+3\mu_2)c_t}{3 \sqrt{3} \mu _2 \left(\left(4 \mu _1-\mu _2\right) \mu _2+4 \mu _1 L^2\right)}, \quad & b_F &= \frac{\left(\mu _2-\mu _1\right) \mu _2}{3 \sqrt{3} \mu _1 L},
\end{aligned}
\end{equation}
and the scalars are
    \begin{align} \label{eq:zerosol4}
        \widetilde{\varphi} = 0, \quad e^{\widetilde{\alpha}} = \sqrt{\frac{\mu_2}{2\mu_1}}, \quad \widetilde{\Sigma}^6 = \frac{\mu_2}{2\mu_1}.
    \end{align}
The solution can be described by two independent parameters as supersymmetry and extremality constrain the solution. More precisely, we choose to use $\mu_1$ and $\mu_2$ as our two parameters, which are related to those in Section~\ref{sec:z=1bhsolutions} via\footnote{These are also the conventions used in \cite{Gutowski:2004yv}.}
\begin{align} \label{eq:mudelta}
    \mu_1 = \frac{2L^2}{e^{4\delta_2}-1}, \qquad \mu_2 = \frac{2L^2}{e^{2\delta_1+2\delta_2}-1}.
\end{align}
We can now proceed to carry out a perturbative expansion with $z=1+\epsilon$ via
	\begin{align} \label{eq:functionsexpansion}
		g_i  = \sum_{n=0}^{n_{\text{max}}} \epsilon^n f^{(n)}, \quad a_{1,2} = \sum_{n=0}^{n_{\text{max}}} \epsilon^n a_{1,2}^{(n)}, \quad W = \sum_{n=0}^{n_{\text{max}}} \epsilon^n W^{(n)},
	\end{align}
where $W = ( \widetilde \alpha, \widetilde \Sigma, \widetilde\varphi )$, and $n_{\text{max}}$ is the number of terms in the expansion. We work up to first order in $\epsilon$, i.e., $n_{\text{max}}=1$.\footnote{As $z$ must satisfy \eqref{eq:zdef}, we assume $\epsilon$ is a small and rational number.} Implementing \eqref{eq:functionsexpansion} is straightforward and the end result is a set of linear algebraic equations that can be solved analytically.

There are eight independent equations of motion as well as the supersymmetric constraint, similar to \eqref{eq:linearconstraintgrav}, which we will discuss in detail in Section~\ref{sec:thermodynamics}. On the other hand, there are eleven parameters to determine, resulting in a solution written with two additional parameters at linear order. As the solution is extremal and supersymmetric, the expected total number of independent parameters must remain two, regardless of the order in $\epsilon$. This indicates that there is freedom in defining the independent parameters.

In principle, the two independent parameters that describe the solution are arbitrary. A convenient choice is to fix two of the integration constants such that the horizon area retains the same form. This effectively means $g_2$ and $g_3$ are not modified at linear order in $\epsilon$, i.e.
\begin{align} \label{eq:gauge}
    g^{(1)}_2 = 0, \quad g^{(1)}_{3} = 0.
\end{align}
This is particularly useful because it leads to the vanishing of corrections to additional thermodynamic quantities, as we shall see in Section~\ref{sec:thermodynamics}. Upon solving the system of linear equations and imposing \eqref{eq:gauge}, the only nonzero corrected constants correspond to the scalars as well as the gauge potential associated with the flavour charges
\begin{equation} \label{eq:firstordersolscalars}
    \begin{aligned}
    \widetilde{\alpha}^{(1)} &= -\frac{2\mu_1 + \mu_2}{16\mu_1}, \quad & \widetilde{\Sigma}^{(1)} &= \frac{\left(\mu _2\right)^{1/6} \left(3 \mu _2-10 \mu _1\right)  }{48 \sqrt[6]{2} \mu _1^{7/6}}, & \quad \widetilde{\varphi}^{(1)} &= -\frac{\left(4 \mu _1^2-2 \mu _2 \mu _1+\mu _2^2\right)}{16 \mu _1^2},
    \end{aligned}
\end{equation}
and
\begin{equation} \label{eq:firstordersolpotential}
    \begin{aligned}
    a^{(1)}_{F} &= \frac{\left(\mu _2-\mu _1\right) L   \sqrt{\left(4 \mu _1-\mu _2\right) \mu _2+4 \mu _1 L^2}}{12 \sqrt{3} \mu _1 \left(\mu _1+2 \mu _2+L^2\right)}, & \quad  b^{(1)}_{F} &= \frac{\left(\mu _1-\mu _2\right) \mu _2}{12 \sqrt{3} \mu _1 L}.
    \end{aligned}
\end{equation}
This suggests that the R-charge, the angular momentum and the associated fugacities along with the entropy will not be corrected at linear order in $\epsilon$ when imposing \eqref{eq:gauge}. Finally, we remark that the hyperscalar is nonzero and cannot be set to zero by fixing the integration constants in another way. This indicates that the hypermultiplet plays an essential role in the general black hole solution.

%%%%%%%%%%%%%%%%%%%%%%%%%%
%%%%%%%%%%%%%%%%%%%%%%%%%%
%%%%%%%%%%%%%%%%%%%%%%%%%%
\subsection{Thermodynamics at the near-horizon} \label{sec:thermodynamics}
%%%%%%%%%%%%%%%%%%%%%%%%%%
%%%%%%%%%%%%%%%%%%%%%%%%%%
%%%%%%%%%%%%%%%%%%%%%%%%%%
To study this black hole solution and its thermodynamics, in particular the entropy, we compute the thermodynamic quantities from the NHEG, up to first-order corrections in $\epsilon$ using the solution \eqref{eq:firstordersolscalars} and \eqref{eq:firstordersolpotential}. We must be more cautious when obtaining the thermodynamic quantities from the horizon, as not all quantities are accessible from this region. Unlike the special $z=1$ case, the lack of global information requires a careful computation of the charges via Komar integrals, ensuring that a Gauss law still holds, e.g., \cite{Bazanski:1990qd,Kastor:2008xb, Kastor:2009wy,Ortin:2021ade,Cano:2023dyg,David:2023gee}. This allows us to choose the horizon as our surface of integration. Komar integrals in the context of NHEGs have been studied recently for AdS$_5$ black holes \cite{Cano:2024tcr}.

Even so, certain quantities are not accessible from this region, in particular the ADM mass, the chemical potentials and the on-shell action. Our primary goal is to compute the entropy, in the microcanonical ensemble -- i.e. as a function of the charges -- to confirm the prediction from the CFT dual. Consequently, the quantities we cannot compute directly, are not essential for this purpose and below we focus on computing the electric charge and the angular momentum.

The presence of the Chern-Simons terms introduces different notions of charge \cite{Marolf:2000cb}. Here we compute the Page charge
\begin{align}
    Q_I &= -\frac{1}{16\pi G_5}\int_{\mathcal{H}} a_{IJ} \star F^{J} + 3 c_{IJK} A^{J} \wedge F^{K}.
\end{align}
which can be shown to satisfy a Gauss law. However, there is still gauge ambiguity which we can fix by demanding regularity of the potential at the horizon. In this way, we obtain a well-defined expression for the electric charge.

If the solution has a Killing vector $\xi$ we have an associated conserved current which on-shell takes the form\footnote{We use the convention of denoting differential forms with boldface.}
\begin{align}
   \mathbf{J}_{\xi} = d \mathbf{Q}_{\xi},
\end{align}
where $\mathbf{Q}_{\xi}$ is the conserved charge associated to $\xi$. The exterior derivative acting on $\mathbf{J}_{\xi}$ is zero which allows us to compute the conserved charge by integrating $\mathbf{J}_{\xi}$ on a spatial hypersurface of the spacetime. Instead, implementing Stoke's theorem allows us to recast the integral as
\begin{align} \label{eq:currentint1}
    Q_{\xi} = \int_{\partial \mathcal{M}} \mathbf{Q}_{\xi},
\end{align}
which must be evaluated at the boundary $\partial\mathcal{M}$ as \eqref{eq:currentint1} does not inherently satisfy a Gauss law. This integration requires knowledge of the asymptotics of the solution, which is not directly accessible from the NHEG. To resolve this, we define a conserved charge $\mathbf{\tilde{Q}}_{\xi} =  \mathbf{Q}_{\xi} - d\mathbf{\Omega}$. where $d\mathbf{\Omega} = \mathbf{J}_{\xi}|_{\text{on-shell}}$. %$d\mathbf{\Omega} = \mathbf{J}_{\xi}|_{\text{on-shell}} = - \frac{1}{2}\star \xi \mathcal{L}$.
The difference between the two charges corresponds to the integral of $\mathbf{\Omega}$ at infinity which vanishes by choosing $\mathbf{\Omega}$ appropriately. With $\mathbf{\Omega}$ fixed, the charge
is now independent of the surface of integration. With this procedure, the angular momentum associated with the Killing vector $\xi_{\phi} = -\partial_{\phi}$ is given by
\begin{align}
    J &= -\frac{1}{16\pi G_5}\int_{\mathcal{H}} \star d \xi + (\xi \cdot A^{I})  \left(a_{IJ} \star F^{J}+2c_{IJK} A^{J} \wedge F^{K} \right),
\end{align}
Imposing the solution up to first order in $\epsilon$ using \eqref{eq:zerosol1}, \eqref{eq:zerosol2}, \eqref{eq:zerosol3}, \eqref{eq:zerosol4}, \eqref{eq:firstordersolscalars} and \eqref{eq:firstordersolpotential}, we find the corrected charges
    \begin{equation} \label{eq:JQRQFlinear}
    \begin{split}
    J &= \frac{2\pi  \mu _2 \left(2 \mu _1 \mu _2+\left(2 \mu _1+\mu _2\right) L^2\right)}{16 G_5 L^3}+\mathcal{O}(\epsilon^2),
    \\
    Q_R &= \frac{\pi  \left(\mu _2 \left(2 \mu _1+\mu _2\right)+2 \left(\mu _1+2 \mu _2\right) L^2\right)}{8 \sqrt{3} G_5 L^2}+\mathcal{O}(\epsilon^2),
    \\
    Q_F &= (4+\epsilon)\frac{\sqrt{3} \pi  \left(\mu _1-\mu _2\right) \left(\mu _2+L^2\right)}{32 G_5 L^2}+\mathcal{O}(\epsilon^2).
    \end{split}
\end{equation}
The leading order term in the expansion of the charges matches that found in the global solution for $z=1$, upon imposing extremality \eqref{eq:extremalhorizon} and supersymmetry \eqref{eq:z=1susyconditionparams} and using the relation \eqref{eq:mudelta}. This confirms that we have indeed chosen a suitable $\mathbf{\Omega}$ and have computed the charges correctly at the horizon. The advantage of the choice \eqref{eq:gauge} is evident due to the simplicity of the corrections of the NHEG solution \eqref{eq:firstordersolscalars} and \eqref{eq:firstordersolpotential} and the corresponding charges \eqref{eq:JQRQFlinear}. Since only the gauge potential associated with the U(1)$_F$ is corrected in \eqref{eq:firstordersolscalars} and \eqref{eq:firstordersolpotential}, only the flavour charge itself is corrected, while $J$ and $Q_R$ remain unchanged. Moreover, as the metric contains no first order corrections in $\epsilon$, the entropy, given by the area formula, remains uncorrected as a function of $\mu_1$ and $\mu_2$ at this order
\begin{align} \label{eq:Slinear}
    S = \frac{\pi ^2 \mu _2 \sqrt{-\mu _2^2+4 \mu _1 \mu _2+4 \mu _1 L^2}}{4 G_5 L}+\mathcal{O}(\epsilon^2).
\end{align}
Before proceeding, we examine two different relations that appear at the NHEG, the first being analogous to the first law of thermodynamics \eqref{eq:firstlawBH}.
To understand why such a differential relation may hold at the horizon, we consider the Maxwell relations, obtained by taking second derivatives of \eqref{eq:firstlawBH} and using the fact that the partial derivatives commute
\begin{align}
    \left(\frac{\partial S}{\partial J}\right)_{T, Q}=\left(\frac{\partial \Omega}{\partial T}\right)_{J, Q_I}, \quad\left(\frac{\partial S}{\partial Q_I}\right)_{T, J}= \left(\frac{\partial \Phi^{I}}{\partial T}\right)_{J, Q_I}.
\end{align}
The right hand side of both equations allows us to make the connection to the fugacities as defined in \eqref{eq:fugacitiesfromfirstlaw}
\begin{equation}\label{eq:fugacitiesfromfirstlaw}
    \omega =  \left(\frac{\partial S}{\partial J}\right)_{T=0, Q}, \qquad \varphi^I =  \left(\frac{\partial S}{\partial Q_I}\right)_{T=0, J}.
\end{equation}
Using \eqref{eq:fugacitiesfromfirstlaw}, we can write down a near-horizon version of the first law, given by
\begin{equation}\label{eq:firstlawhorizon}
    dS=\omega_{}dJ_{}+\varphi^I dQ_I.
\end{equation}
This clarifies how these quantities can be computed, $\omega$ from the fibration in the metric and $\varphi^{I}$ from the regularity of the potential at the horizon
\begin{equation}
    \omega = 2\pi \tilde{\omega}, \quad \varphi^{R,F} = 2\pi a_{R,F}.
\end{equation}
Their explicit expressions up to first order in $\epsilon$ are
\begin{equation}  \label{eq:fugacitieslinear}
    \begin{split}
    \omega &=-\frac{2\pi\left(2 \mu _1+\mu _2\right) L^2}{\left(\mu _1+2 \mu _2+L^2\right) \sqrt{\left(4 \mu _1-\mu _2\right) \mu _2+4 \mu _1 L^2}} + \mathcal{O}(\epsilon^2),
    \\
    \varphi ^R &= -\frac{2\pi\left(2 \mu _1+\mu _2\right) L^3}{\sqrt{3} \left(\mu _1+2 \mu _2+L^2\right) \sqrt{\left(4 \mu _1-\mu _2\right) \mu _2+4 \mu _1 L^2}}+ \mathcal{O}(\epsilon^2),
    \\
    \varphi ^F &= (4-\epsilon)\frac{2\pi \left(\mu _1-\mu _2\right) L\left(3 \mu _2+2 L^2\right)}{6 \sqrt{3} \left(\mu _1+2 \mu _2+L^2\right) \sqrt{\left(4 \mu _1-\mu _2\right) \mu _2+4 \mu _1
   L^2}}+ \mathcal{O}(\epsilon^2).
    \end{split}
\end{equation}
As seen from \eqref{eq:firstordersolscalars} and \eqref{eq:firstordersolpotential}, only the fugacity associated to the flavour charge receives corrections at first order.

The approach often taken for obtaining a well-defined quantum statistical relation involves first imposing the supersymmetric limit via \eqref{eq:z=1susyconditioncharges}, followed by extremality. This results in complex charges and fugacities. While the charges become real upon extremality, the fugacities remain complex and are related by a supersymmetric linear constraint such as \eqref{eq:linearconstraintgrav}. In constrast, here we have taken a different procedure, where we start with the NHEG geometry, i.e an extremal solution, and only then impose the supersymmetric limit. This reverse order of limits results in a real linear supersymmetry constraint
\begin{equation} \label{eq:NHEGsusyconstraint}
     \omega -\frac{\sqrt{3}}{L} \varphi^{R}=0.        
\end{equation}
Substituting the solution and the thermodynamic quantities \eqref{eq:JQRQFlinear}, \eqref{eq:Slinear} and \eqref{eq:fugacitieslinear} up to first order in $\epsilon$, we find both \eqref{eq:firstlawhorizon} and \eqref{eq:NHEGsusyconstraint} are satisfied. 

With the thermodynamic charges \eqref{eq:JQRQFlinear} and entropy \eqref{eq:Slinear} at hand, we now set out to confirm the prediction of the microstate counting \eqref{eq:fullentropy} from the dual CFT. Since the Bekenstein-Hawking entropy expressed in terms of $\mu_1$ and $\mu_2$ in \eqref{eq:Slinear} does not receive a first order correction in $\epsilon$, we can use \eqref{eq:z=1entropygravity} as a guide to write down the entropy as a function of the charges%, we can write down an expression for $S$ as a function of the charges, guided by \eqref{eq:z=1entropygravity},
\begin{equation}\label{eq:correctedentropy}
    S = \frac{2\pi}{\sqrt{3}}\sqrt{-\frac{3\pi L^3 J}{2G_5} - \frac{L^2}{12}(4-\epsilon)^2 Q_F^2 + 3 L^2 Q_R^2} + \mathcal{O}(\epsilon^2),
\end{equation} 
which holds to first order in $\epsilon$. On the other hand, we can start from the CFT prediction for the entropy \eqref{eq:fullentropy} and translate to the gravity charges using the dictionary \eqref{eq:gravityCFTrelation} and
\begin{equation}
    G_5 = \frac{3\pi L^3}{64 e^{3f_0+2g_0} (\mathfrak{g}-1) N^3} = \frac{3L^3 \pi(4-\epsilon)}{32(\mathfrak{g}-1)N^3} + \mathcal{O}(\epsilon^2).
\end{equation}
We find that this matches precisely \eqref{eq:correctedentropy}, therefore indicating a highly nontrivial check that the NHEG solution corresponds to a black hole geometry that dominates the gravitational path integral, as predicted from the CFT index.

Although the solution is currently limited to first order in $\epsilon$, the result suggests that the perturbative expansion in $\epsilon$ to higher orders should provide additional consistency checks. This is a next natural step and one may then hope to be able to sum the series to find the complete solution. We comment on this further in Section~\ref{section:conclusions}.

%%%%%%%%%%%%%%%%%%%%%%%%%%%%%%
\section{Final comments and open questions} \label{section:conclusions}
%%%%%%%%%%%%%%%%%%%%%%%%%%%%%%

We have constructed novel black hole solutions in a family of 5d $\mathcal{N}=2$ gauged supergravities coupled to two vector multiplets and a universal hypermultiplet. These black holes have equal angular momenta $J$ and electric charges $Q_R$ and $Q_F$. We first considered the special point $z=1$, from which the solution at $z=-1$ follows straightforwardly. In this limit, the hypermultiplet vanishes and as a result the theory can be viewed as a subsector of the STU model. This allowed us to identify a black hole solution using the previously known solutions of  \cite{Gutowski:2004yv,Cvetic:2004ny,Cvetic:2005zi,Kunduri:2006ek}.  We followed the usual holographic analysis to derive the on-shell action and the entropy in the supersymmetric and extremal limit. This successfully matched the CFT prediction where we utilized the identity \eqref{eq:trick} to aid us in performing the Legendre transform of the index.

Outside the special case of $z=\pm1$, there is no longer a closure relation and the identity \eqref{eq:trick} does not hold. Nevertheless, we identified an analogous relation, allowing us to directly obtain the entropy, whose form is now much more involved. We then constructed the NHEG perturbatively close to $z=1$ and upon a suitable choice of integration constants \eqref{eq:gauge}, showed that the metric does not receive corrections at first order. The Bekenstein-Hawking entropy, written in terms of the charges or in our particular choice of parametrization, confirms the prediction arising from the CFT index.

We have focused solely on the first term in the perturbative expansion around $z=1$ at the NHEG. It is possible to solve higher orders in the series expansion, with each subsequent order depending on the solutions of the previous order, forming a recursive relation. By continuing this process, one can perhaps extrapolate the series and determine the full function. As in the first order, we may consider a similar condition to \eqref{eq:gauge} at higher orders to eliminate two of the metric corrections. This proved to be particularly useful at first order, as it also led to additional cancellations, leaving only the corrections to the scalars and the gauge potential associated with the flavor charge. As shown, this leads to very simple corrections to the flavour charge and associated fugacity. If this structure of corrections still holds at higher orders, determining the corrections at arbitrary values of $z$ would become more tractable.

What remains is to construct the full solution, whether perturbatively around $z=1$ or for general $z$. It is possible that the analytic solution is out of reach and in this case, one can proceed with numerical methods where the expansion around the horizon and at infinity can be matched in an intermediate value of the radial coordinate. This would be implemented without taking the near-horizon extremal limit, as was done in Section~\ref{sec:newbhsolutions}, as the change in coordinate system makes it difficult to compare the NHEG to the asymptotic expansion. A final tool that could be used in the search for the global solution are the supersymmetry variation equations. As these are first order differential equations, these might be more easily solved than the second order equations of motion. The challenge here is identifying the projectors that must be imposed on the Killing spinors. In either case, the full geometry would give us access to the global properties of the solution, which includes the ADM mass and more importantly the on-shell action, providing a more complete holographic match. This is left for future work.

It would also be interesting to go beyond leading order in the large $N$ expansion. This would involve studying the contribution to the index from the linear 't Hooft anomalies, which has not yet been well understood from a holographic point of view. On the gravity side, this would correspond to adding higher derivative corrections to the theory, in the spirit of \cite{Baggio:2014hua}. At present, these corrections have not been constructed and one would take a pragmatic approach of using the off-shell formalism to find the corrections. However, the off-shell supersymmetric transformations for the hypermultiplets require an infinite number of auxiliary fields \cite{deWit:1984rvr} and therefore, computing the four-derivative corrections is out of reach with the current tools at hand.

Regardless, we can still try to derive a prediction for the subleading corrections to the entropy from the field theory dual, taking a similar approach and manipulating the extremization equations arising from the Legendre transform. It is, however, unclear if this manipulation can lead to a single equation for $\Lambda$ that determines the entropy since the sixth-order relation \eqref{eq:newtrick} would no longer hold. One would have to look for a similar equation of higher order, which is not guaranteed to exist. One may take the approach of \cite{Cassani:2019mms, Cassani:2024tvk} and make certain assumptions on the leading and subleading order behavior of the anomalies. Even if an equation does exists, it does not necessarily mean that the resulting solutions for $\Lambda$ can be solved analytically.

Along those lines, a key open question is to determine a complete generalization for performing the Legendre transform for 4d $\mathcal{N}=1$ SCFTs. It was shown in e.g., \cite{Amariti:2019mgp} that there sometimes exists an identity similar to \eqref{eq:closurerelation} even when there a closure relation is not satisfied. We may ask what properties of the CFT accounts for this, in particular when the dual gravity theory is not known. There might be certain geometrical properties on either side of the duality that determine whether an identity can be found to simplify the extremization procedure and predict the order of the polynomial. We plan to investigate this further in future work.

\section*{Acknowledgments}
We are grateful for Pablo Cano, Alfredo Gonz\'{a}lez Lezcano, Joel Karlsson, Daniel Mayerson, Hynek Paul, Enrico Turetta, Jesse van Muiden and Valentin Reys for useful discussions. We thank Nikolay Bobev and Vasil Dimitrov for the early stages of this collaboration. We greatly appreciate Pablo Cano and Alfredo Gonz\'{a}lez Lezcano for comments on earlier versions of this manuscript. This research is supported in part by the Odysseus grant (G0F9516N Odysseus). MD is supported by the Postdoctoral Fellows of the Research Foundation - Flanders with the grant 1235324N and AV by the PhD Fellowship of the Research Foundation - Flanders with grant 1102724N.
	
\appendix

%%%%%%%%%%%%%%%%%%%%%%%%%%%%%%
\section{Sixth order equation for \texorpdfstring{$\Lambda$}{} in terms of the charges} \label{app:sexticeq}
%%%%%%%%%%%%%%%%%%%%%%%%%%%%%%
The Legendre transform of the index \eqref{eq:CFTindexleadingorder} leads to a polynomial equation for $\Lambda$ given in \eqref{eq:sexticeq}. This is written in terms of $q_i$ which are functions of the angular momenta $\hat{J}_{1,2}$ and charges $\hat{Q}_{R,F}$ of the black hole, as well as the numbers $(N,\mathfrak{g},z)$ that are contained in the anomalies $k_{IJK}$. We present here the full expressions for the $q_i$.
\begin{align} \label{eq:fullq's}
    k_{FFF}^2 q_0 
    &= k_{RRR} (k_{RFF} (6 \hat{Q}_F^4 \hat{Q}_R^2-\tfrac{27}{8} \hat{J}_1 \hat{J}_2 \hat{Q}_F k_{FFF} \hat{Q}_R^2)-2 \hat{Q}_F^3 k_{FFF} \hat{Q}_R^3\nonumber
    \\&
    +\tfrac{27}{4} \hat{J}_1 \hat{J}_2 \hat{Q}_F^2 k_{RFF}^2 \hat{Q}_R  +\tfrac{9}{8} \hat{J}_1 \hat{J}_2 k_{FFF}^2 \hat{Q}_R^3+\tfrac{81}{64} \hat{J}_1^2 \hat{J}_2^2 k_{RFF}^3)+k_{RRR}^2 (\tfrac{9}{8} \hat{J}_1 \hat{J}_2 \hat{Q}_F^3 k_{FFF}\nonumber
    \\&
    +\hat{Q}_F^6+\tfrac{81}{256} \hat{J}_1^2 \hat{J}_2^2 k_{FFF}^2) -6 \hat{Q}_F k_{FFF} k_{RFF} \hat{Q}_R^5+9 \hat{Q}_F^2 k_{RFF}^2 \hat{Q}_R^4+k_{FFF}^2 \hat{Q}_R^6 \nonumber
    \\&
    +\tfrac{9}{4} \hat{J}_1 \hat{J}_2 k_{RFF}^3 \hat{Q}_R^3, \nonumber
    \\ \nonumber \\
    k_{FFF}^2 q_1 \nonumber 
    &=
    k_{RRR} (k_{RFF} (-\tfrac{27}{8} \hat{Q}_F k_{FFF} \hat{Q}_R (\hat{J}_2 \hat{Q}_R+\hat{J}_1 (\hat{Q}_R-2 \hat{J}_2))-12 \hat{Q}_F^4 \hat{Q}_R)\nonumber
    \\&
    +6 \hat{Q}_F^3 k_{FFF} \hat{Q}_R^2 -\tfrac{27}{4} \hat{Q}_F^2 k_{RFF}^2 (\hat{J}_1 (\hat{J}_2-\hat{Q}_R)-\hat{J}_2 \hat{Q}_R)+\tfrac{9}{8} k_{FFF}^2 \hat{Q}_R^2 (\hat{J}_2 \hat{Q}_R \nonumber
    \\&
    +\hat{J}_1 (\hat{Q}_R-3 \hat{J}_2))+\tfrac{81}{32} \hat{J}_1 \hat{J}_2 (\hat{J}_1+\hat{J}_2) k_{RFF}^3)+k_{RRR}^2 (\tfrac{9}{8} (\hat{J}_1+\hat{J}_2) \hat{Q}_F^3 k_{FFF} \nonumber
    \\&
    +\tfrac{81}{128} \hat{J}_1 \hat{J}_2 (\hat{J}_1+\hat{J}_2) k_{FFF}^2)+30 \hat{Q}_F k_{FFF} k_{RFF} \hat{Q}_R^4-36 \hat{Q}_F^2 k_{RFF}^2 \hat{Q}_R^3 \nonumber 
    \\&
    -6 k_{FFF}^2 \hat{Q}_R^5+\tfrac{9}{4} k_{RFF}^3 \hat{Q}_R^2 ((\hat{J}_1+\hat{J}_2) \hat{Q}_R-3 \hat{J}_1 \hat{J}_2), \nonumber 
    \\ \nonumber \\
    k_{FFF}^2 q_2 
    &=
    k_{RRR} (k_{RFF} (6 \hat{Q}_F^4-\tfrac{27}{8} \hat{Q}_F k_{FFF} (-2 (\hat{J}_1+\hat{J}_2 \hat{Q}_R+\hat{J}_1 \hat{J}_2+\hat{Q}_R^2-6 \hat{Q}_F^3 k_{FFF} \hat{Q}_R \nonumber
    \\&
    -\tfrac{27}{4} \hat{Q}_F^2 k_{RFF}^2 (\hat{J}_1+\hat{J}_2-\hat{Q}_R+\tfrac{9}{8} k_{FFF}^2 \hat{Q}_R (-3 (\hat{J}_1+\hat{J}_2 \hat{Q}_R+3 \hat{J}_1 \hat{J}_2+\hat{Q}_R^2+\tfrac{81}{64} (\hat{J}_1^2 \nonumber
    \\& 
    +4 \hat{J}_2 \hat{J}_1+\hat{J}_2^2 k_{RFF}^3+k_{RRR}^2 (\tfrac{9}{8} \hat{Q}_F^3 k_{FFF}+\tfrac{81}{256} (\hat{J}_1^2+4 \hat{J}_2 \hat{J}_1+\hat{J}_2^2 k_{FFF}^2 \nonumber
    \\& -60 \hat{Q}_F k_{FFF} k_{RFF} \hat{Q}_R^3 +54 \hat{Q}_F^2 k_{RFF}^2 \hat{Q}_R^2+15 k_{FFF}^2 \hat{Q}_R^4 \nonumber
    \\&
    +\tfrac{9}{4} k_{RFF}^3 \hat{Q}_R (-3 (\hat{J}_1+\hat{J}_2 \hat{Q}_R+3 \hat{J}_1 \hat{J}_2+\hat{Q}_R^2), \nonumber
    \\ \nonumber \\ \nonumber
    k_{FFF}^2 q_3 
    &=
    k_{RRR} (-\tfrac{27}{8} \hat{Q}_F k_{FFF} k_{RFF} (\hat{J}_1+\hat{J}_2-2 \hat{Q}_R)+2 \hat{Q}_F^3 k_{FFF}-\tfrac{27}{4} \hat{Q}_F^2 k_{RFF}^2 \nonumber
    \\&
    -\tfrac{9}{8} k_{FFF}^2 (3 \hat{Q}_R (\hat{Q}_R-\hat{J}_2)+\hat{J}_1 (\hat{J}_2-3 \hat{Q}_R))+\tfrac{81}{32} (\hat{J}_1+\hat{J}_2) k_{RFF}^3) \nonumber
    \\&
    +60 \hat{Q}_F k_{FFF} k_{RFF} \hat{Q}_R^2-36 \hat{Q}_F^2 k_{RFF}^2 \hat{Q}_R+\tfrac{81}{128} (\hat{J}_1+\hat{J}_2) k_{FFF}^2 k_{RRR}^2-20 k_{FFF}^2 \hat{Q}_R^3 \nonumber
    \\&
    -\tfrac{1}{4} 9 k_{RFF}^3 (3 \hat{Q}_R (\hat{Q}_R-\hat{J}_2)+\hat{J}_1 (\hat{J}_2-3 \hat{Q}_R)),
    \nonumber \\ \nonumber \\
    k_{FFF}^2 q_4 &=
    k_{RRR} (-\tfrac{27}{8} \hat{Q}_F k_{FFF} k_{RFF}-\tfrac{9}{8} k_{FFF}^2 (\hat{J}_1+\hat{J}_2-3 \hat{Q}_R)+\tfrac{81 k_{RFF}^3}{64}) \nonumber
    \\&
    -30 \hat{Q}_F k_{FFF} k_{RFF} \hat{Q}_R+9 \hat{Q}_F^2 k_{RFF}^2+15 k_{FFF}^2 \hat{Q}_R^2+\tfrac{81}{256} k_{FFF}^2 k_{RRR}^2 \nonumber
    \\&
    -\tfrac{1}{4} 9 k_{RFF}^3 (\hat{J}_1+\hat{J}_2-3 \hat{Q}_R), \nonumber
    \\ \nonumber \\
    k_{FFF}^2 q_5 &= 6 \hat{Q}_F k_{FFF} k_{RFF}-6 k_{FFF}^2 \hat{Q}_R-\tfrac{9}{8} k_{FFF}^2 k_{RRR}-\tfrac{1}{4} 9 k_{RFF}^3.
\end{align}

%%%%%%%%%%%%%%%%%%%%%%%%%%%%%%
\section{Consistent truncation to 7d} \label{app: consistent truncation to 7d}
%%%%%%%%%%%%%%%%%%%%%%%%%%%%%%
In this appendix, we review the dimensional reduction of 11d supergravity on a squashed four-sphere to the 7d U(1)$\times$U(1) gauged supergravity of \eqref{eq:7DSUGRAaction1}. A well-known reduction brings the 11d theory down to 7d $\mathcal{N}=4$ $SO(5)$ gauged supergravity \cite{Pernici:1984xx,Nastase:1999cb,Donos:2010ax}. One can then use an additional consistent truncation to further reduce to the U(1)$\times$U(1) invariant subsector of this 7d theory, which contains a metric $ds_7{}^2$, a 3-form $\mathcal{S}$, two Abelian gauge fields $\mathcal{A}_{A,B}$ with field strengths $\mathcal{F}_{A,B}$ and two scalars $\lambda_{A,B}$ \cite{Liu:1999ai}. Here we present the direct reduction formulae from 11d supergravity to the 7d U(1)$\times$U(1) invariant theory. Starting from the action of 11d supergravity
\begin{equation}\label{eq:11dsugraaction}
    I_{11} = \frac{1}{16 \pi G_{11}}\int \qty(  \star_{11} R_{11} - \frac{1}{2} F^{(4)} \wedge \star_{11} F^{(4)} + \frac{1}{6}F^{(4)} \wedge F^{(4)} \wedge A^{(3)} ) \,,
\end{equation}
one chooses an ansatz for the 11d metric that takes on the form of a warped product
\begin{equation}
    \dd{s}^2_{11} = \Delta^{1/3} \dd{s}^2_7 + \Delta^{-2/3} \dd{s}^2_4 \,.
\end{equation}
The $S^4$ can be parametrized by four angles $ (\alpha, \, \beta, \, \phi_A, \, \phi_B)$, of which $ \phi_{A,B}$ parametrize the U(1)$\times$U(1) isometry directions and have covariant derivatives\footnote{Here $m$ is an inverse length scale.} $ D\phi_{A,B} = \dd{\phi_{A,B}} + m \mathcal{A}_{A,B}$. The other two angles $(\alpha,\beta)$ are better expressed in terms of alternative, constrained coordinates $ \nu_{1,2,3}$ as
\begin{align}
    \nu_1 & = \sin \alpha \cos \beta \,, \quad \nu_2 = \sin \alpha \sin \beta \,, \quad \nu_3 = \cos \alpha \,;  \qquad \nu_1^2 + \nu_2^3 + \nu_3^2 = 1 \,.
\end{align}
In terms of $D\phi_{A,B}$, the $\nu_i$ and two 7d scalars $\lambda_{A,B}$, the metric on the $S^4$ is given by
\begin{align}\label{eq:S4metric}
    \dd{s}^2_4 & = \frac{1}{m^2} \qty( e^{-2\lambda_A} \qty[ \dd{\nu_1}^2 + \nu_1^2 (D \phi_A)^2 ] + e^{-2\lambda_B} \qty[ \dd{\nu_2}^2 + \nu_2^2 (D \phi_B)^2 ] + e^{4\lambda_A + 4\lambda_B} \dd{\nu_3}^2) \,.
\end{align}
The 4-form flux is expressed in terms of the 7d fields as
\begin{align}
    F^{(4)} & = - e^{-4\lambda_A-4\lambda_B} \nu_3 \star_7 \mathcal{\mathcal{S}} + \frac{1}{m} \mathcal{S} \wedge \dd{\nu_3} \nonumber \\ 
    & \quad + \frac{1}{\Delta m^2} \Bigg[ - e^{2\lambda_A} \nu_1^2 \, D \phi_A \wedge \mathcal{F}_B \wedge \dd{\nu_{3}} - e^{2\lambda_B} \nu_2^2 \, D \phi_B \wedge \mathcal{F}_A \wedge \dd{\nu_{3}} \nonumber \\
    & \qquad \qquad \qquad \, + e^{-4\lambda_A-4\lambda_B} \nu_1 \nu_3 \, D \phi_A \wedge \mathcal{F}_B \wedge \dd{\nu_1} + e^{-4\lambda_A-4\lambda_B} \nu_2 \nu_3 \, D \phi_B \wedge \mathcal{F}_A \wedge \dd{\nu_2} \Bigg] \nonumber \\
    & \quad + \frac{U \nu_1 \nu_2}{\Delta^2m^3} \Bigg[ \nu_1 \, D \phi_A \wedge D \phi_{B} \wedge \dd{\nu_2} \wedge \dd{\nu_{3}} + \nu_2 \, D \phi_B \wedge D \phi_A \wedge \dd{\nu_1} \wedge \dd{\nu_{3}} \nonumber \\
    & \qquad \qquad \qquad + \nu_3 \, D \phi_A \wedge D \phi_{B} \wedge \dd{\nu_1} \wedge  \dd{\nu_2} \Bigg]  \nonumber\\
    & \quad + \frac{\nu_1\nu_2}{\Delta^2m^3} \Bigg[ 2 e^{2\lambda_A + 2\lambda_B} \nu_1 \nu_2 \, D \phi_A \wedge D \phi_B \wedge \dd{\qty(\lambda_A - \lambda_B)} \wedge \dd{\nu_3}  \nonumber \\
    & \qquad \qquad \qquad + 2 e^{-2\lambda_A - 4\lambda_B} \nu_1 \nu_3 \, D \phi_A \wedge D \phi_B \wedge \dd{(3\lambda_A - 2\lambda_B)} \wedge \dd{\nu_2} \nonumber\\
    & \qquad \qquad \qquad + 2 e^{-4\lambda_A - 2\lambda_B} \nu_2 \nu_3 \, D \phi_A \wedge D \phi_B \wedge \dd{(3\lambda_A + 2\lambda_B)} \wedge \dd{\nu_1} \Bigg] \,,
\end{align}
where $\mathcal{F}_{A,B} = \dd{\mathcal{A}_{A,B}}$ and $\Delta$ and $U$ are given by
\begin{align}
    \begin{aligned}
        \Delta & = e^{2\lambda_A} \nu_1^2 + e^{2\lambda_B} \nu_2^2 + e^{-4\lambda_A - 4 \lambda_B} \nu_3^2 \,, \\
        U & = -\qty(2e^{2\lambda_A + 2\lambda_B} + e^{-2\lambda_A - 4\lambda_B}) \nu_1^2 -(2e^{2\lambda_A + 2\lambda_B} + e^{-4\lambda_A - 2\lambda_B}) \nu_2^2  \\
        & \qquad + \qty(e^{-8\lambda_A - 8\lambda_B} - 2e^{-2\lambda_A - 4\lambda_B} - 2 e^{-4\lambda_A - 2\lambda_B}) \nu_3^2 \,.
    \end{aligned}
\end{align}
After implementing the expressions above in \eqref{eq:11dsugraaction}, the resulting 7d theory has the following bosonic action
\begin{equation}
    \begin{split}
        I_{7} & = \frac{1}{16\pi G_7}\int \Bigg[  \star_7 (R_7 - V_7) - 5 \dd{\lambda_+}\wedge \star_7 \dd{\lambda_+} - \dd{\lambda_-}\wedge \star_7 \dd{\lambda_-} - \frac{1}{2} e^{-4\lambda_A}\mathcal{F}_A\wedge \star_7 \mathcal{F}_A\\
        &\qquad \qquad \quad - \frac{1}{2} e^{-4\lambda_B}\mathcal{F}_B\wedge \star_7 \mathcal{F}_B - \frac{1}{2} e^{-4\lambda_+} \mathcal{S}\wedge \star_7 \mathcal{S} + \frac{1}{2m} \, \mathcal{S}\wedge \dd{\mathcal{S}}\\
        &\qquad \qquad \quad - \frac{1}{m} \mathcal{F}_A\wedge \mathcal{F}_B \wedge \mathcal{S} + \frac{1}{4m}\qty(\mathcal{A}_A\wedge \mathcal{F}_A\wedge \mathcal{F}_B\wedge \mathcal{F}_B + \mathcal{A}_B \wedge \mathcal{F}_B \wedge \mathcal{F}_A \wedge \mathcal{F}_A) \Bigg],
    \end{split}
\end{equation}
where we have defined $ \lambda_\pm = \lambda_A \pm \lambda_B$ and $ V$ is the potential for the scalars, given by
\begin{equation}
    V_7 = \frac{m^2}{2} \left(e^{-8 \lambda_A-8 \lambda_B}-4 e^{-2 \lambda_A-4 \lambda_B}-4 e^{-4 \lambda_A-2 \lambda_B}-8 e^{2 \lambda_A + 2 \lambda_B}\right).
\end{equation}
The equations of motion of this theory are as follows
\begin{align} \label{eq:7deom}
    \begin{aligned}
        d\mathcal{S} & = m \, e^{-4\lambda_+} \star_7 \mathcal{S} + \mathcal{F}_A \wedge \mathcal{F}_B \,, \\
        d\left(e^{-4\lambda_{A,B}}  \star_7 \mathcal{F}_{A,B}\right) & = \frac{1}{m} \mathcal{F}_{A,B} \wedge \mathcal{F}_{B,A} \wedge \mathcal{F}_{B,A} - \frac{1}{m} \mathcal{F}_{B,A} \wedge d\mathcal{S} \,, \\
        d \star_7 d\left(3 \lambda_{A,B} + 2 \lambda_{B,A}\right) & = - \frac{1}{2} e^{-4\lambda_{A,B}} \mathcal{F}_{A,B} \wedge  \star_{7} \mathcal{F}_{A,B} - \frac{1}{2} e^{-4\lambda_+} \mathcal{S} \wedge  \star_7 \mathcal{S} + \frac{1}{4} \star_7 \pdv{V_7}{\lambda_{A,B}} \,, \\
        R_{MN} & = 5 \partial_M \lambda_+ \partial_N \lambda_- + \partial_M \lambda_- \partial_N \lambda_+ + \frac{1}{2} e^{-4\lambda_A} \mathcal{F}_{A \, MN}^{2} + \frac{1}{2} e^{-4\lambda_B} \mathcal{F}_{B \, MN}^{2} \\
        + \frac{1}{4} & e^{-4 \lambda_+} \mathcal{S}^2_{MN} + \frac{1}{5}g_{MN}\left[V_7 - \frac{1}{4} e^{-4\lambda_A} \mathcal{F}_A^2 - \frac{1}{4} e^{-4\lambda_B} \mathcal{F}_A^2 - \frac{1}{6} e^{-4\lambda_+} \mathcal{S}^2\right] \,,
    \end{aligned}
\end{align}
where $M,N,\ldots$ are curved 7d indices and we have defined $\mathcal{F}_{A,B \, MN}^2 = \mathcal{F}\indices{_{A,B \, M}^R} \mathcal{F}_{A,B \, N R}$, $\mathcal{S}_{MN}^2 = \mathcal{S}\indices{_M^{RS}}\mathcal{S}_{NRS}$, $\mathcal{F}_{A,B}^2 = \mathcal{F}_{A,B}^{MN}\mathcal{F}_{A,B \, MN}$, 
$\mathcal{S}^2 = \mathcal{S}^{MNR}\mathcal{S}_{MNR}$.

%%%%%%%%%%%%%%%%%%%%%%%%%%%%%%
\section{Five-dimensional \texorpdfstring{$\mathcal{N}=2$}{} gauged supergravity}\label{app:multipletconventions}
%%%%%%%%%%%%%%%%%%%%%%%%%%%%%%
Here we present some more details on our conventions in \eqref{eq:5dactionnotheta1} and the multiplet structure of matter-coupled five-dimensional $\mathcal{N}=2$ gauged supergravity. In general, the bosonic sector of 5d $\mathcal{N}=2$ supergravity coupled to $n_V$ vector multiplets and $n_H$ hypermultiplets has the following action \cite{Gunaydin:1999zx,Ceresole:2000jd,Bergshoeff:2004kh}
\begin{equation}
    \begin{split}
        I_5 = \frac{1}{16\pi G_5}\int \left[ \right. & \left. (R_5 -2V_5 )\star_5 1 - \frac{1}{2}g_{xy}\D \phi^x \wedge \star_5 \D \phi^y - \frac{1}{2}g_{XY}\D q^X \wedge\star_5 \D q^Y \right.\\
        &\left.- \frac{3}{4}a_{ij} F^i \wedge\star_5 F^j + \frac{1}{2}c_{ijk}A^i \wedge F^j \wedge F^k\right],
    \end{split}
\end{equation}
including a metric $g_{\mu\nu}$, $n_V+1$ vectors\footnote{With $i$ running over $0,\ldots, n_V$.} $A^i_\mu$, $n_V$ vector multiplet scalars $\phi^x$ and $4n_H$ hyperscalars $q^X$\footnote{We use lower case indices ${}^x$ for the vector multiplets and upper case indices ${}^X$ for the hypermultiplets.}. The vector multiplet scalars parametrize a very special manifold with metric $g_{xy}$ which also determines $a_{ij}$ and the hyperscalars parametrize a quaternionic manifold with metric $g_{XY}$. The set of real constants $c_{ijk}$ are symmetric under permutations of $(ijk)$. One can reparametrize the vector multiplet scalars from $n_V$ scalars $\phi^x$ to $n_V+1$ scalars $h^i$ with a constraint
\begin{equation}
    c_{ijk}h^i h^j h^k = 1.
\end{equation}
The metric on the vector moduli space is then given by
\begin{equation}
    g_{xy} = \frac{3}{2}\partial_x h^i \partial_y h^j a_{ij},
\end{equation}
and the covariant derivatives of the vector multiplet scalars can be rewritten as
\begin{equation}
    \D_\mu h^i = \partial_x h^i \D_\mu \phi^x.
\end{equation}
The action is then
\begin{equation}
    \begin{split}
        I_5 = \frac{1}{16\pi G_5}\int \left[ (R_5 \right.& \left. -2V_5)\star_5 1 - \frac{3}{2}a_{ij}\D h^i \wedge \star_5 \D h^j - g_{XY}\D q^X \wedge\star_5 \D q^Y \right.\\
        &\left.- \frac{3}{2}a_{ij} F^i \wedge \star_5 F^j -\frac{1}{2} c_{ijk}A^i \wedge F^j \wedge F^k \right].
    \end{split}
\end{equation}
We now set $n_V=2$ and $n_H=1$ and choose the moduli space of the scalars to be
\begin{equation}
    \mathcal{M} = \mathcal{M}_V \times \mathcal{M}_H = \left(\mathbb{R}^+ \times \text{SO}(1,1)\right)\times\frac{\text{SU}(2,1)}{\text{SU}(2)_H \times \text{U}(1)}.
\end{equation}
The only non-trivial covariant derivatives are those for the hypermultiplet scalars, and the nonzero components of $c_{ijk}$ are $c_{0IJ}=c_{I0J} = c_{IJ0} = \frac{1}{3}\eta_{IJ}$ with $I,J=1,2$. We redefine
\begin{equation}
    h^0 = \Sigma^{-2}, \qquad h^I = -\Sigma H^I,
\end{equation}
 with $\Sigma$ representing the $\mathbb{R}^+$ and the $H^I$ parametrizing the $SO(1,1)$ by satisfying the constraint
\begin{equation}
    -(H^1)^2 + (H^2)^2 = 1.
\end{equation}
The metric $a_{ij}$ for these scalars is
\begin{equation}
    \begin{split}
        a_{00} = \frac{1}{3}\Sigma^4, \qquad a_{0I} = 0, \qquad 
        a_{I J} = \frac{2}{3}\Sigma^{-2}\begin{pmatrix}
        &2(H^1)^2 + 1 &-2H^1H^2\\
        &-2H^1H^2 &2(H^2)^2 - 1
    \end{pmatrix}.
    \end{split}
\end{equation}
The action is then given by
\begin{equation}\label{eq:CJPWaction}
    \begin{split}
        I_5 = &\frac{1}{16\pi G_5}\int \left[ (R_5-2 V_5)\star_51 - \frac{1}{2}\Sigma^4 F^0 \wedge \star_5 F^0 - \frac{3}{2}a_{I J}F^I \wedge\star_5 F^J - 2\Sigma^{-2}d\Sigma\wedge\star_5 d\Sigma \right.\\
        &\left. - \frac{3}{2} a_{I J}d(\Sigma H^I)\wedge\star_5 d(\Sigma H^J) - g_{X Y}\mathcal{D}q^X\wedge\star_5\mathcal{D}q^Y - A^0\wedge(F^1\wedge F^1 - F^2\wedge F^2) \right].
    \end{split}
\end{equation}
We choose $q^X = (\varphi,\xi,\theta_1,\theta_2)$ such that the metric on $\mathcal{M}_H$ is
\begin{equation}
    g_{XY} dq^X dq^Y = 2d\varphi^2 + e^{2\varphi}(d\theta_1^2 + d\theta_2^2) + \frac{1}{2}e^{4\varphi}(d\xi - \theta_1d\theta_2 + \theta_2d\theta_1)^2.
\end{equation}
The covariant derivatives of the hypers are
\begin{equation}
    \begin{split}
        \mathcal{D}\varphi = d\varphi,\quad
        \mathcal{D}\xi = d\xi + \frac{1}{R}A^0 - \frac{1}{R}(z A^1 - A^2),\quad
        \mathcal{D}\theta_1 = d\theta_1 + \frac{2}{R}A^2 \theta_2,\quad
        \mathcal{D}\theta_2 = d\theta_2 - \frac{2}{R}A^2 \theta_1.
    \end{split}
\end{equation}
This theory contains three abelian gauge fields\footnote{$n_V + 1 = 3$} $A^{0,1,2}$, although on-shell not all three are independent since one linear combination is massive and will be set to zero. This massive gauge field is determined by the covariant derivative for the Stueckelberg scalar
\begin{equation}
    \mathcal{D}\xi=  d\xi + \frac{1}{R}A^0 - \frac{1}{R}(z A^1 - A^2),
\end{equation}
therefore we will set $A^0 = (zA^1 - A^2)$ on shell. Once the massive gauge field has been eliminated, the Stueckelberg scalar $\xi$ decouples and can be discarded. Finally, we parametrize $H^{1,2}$ by one scalar $\alpha$ as
\begin{equation}
    H^1 = \sinh \alpha, \quad H^2 = \cosh \alpha.
\end{equation}
With the above substitutions, we can simplify the action to find
\begin{equation}\label{eq:5dactionsimplified}
    \begin{split}
        I_5 = &\frac{1}{16\pi G_5}\int \left[ (R_5-2 V_5)\star_51 -\frac{3}{\Sigma^2}d\Sigma\wedge\star_5 d\Sigma - d\alpha\wedge\star_5 d\alpha - 2 d\varphi\wedge\star_5 d\varphi \right.\\
        & + e^{2\varphi}(\D\theta_1\wedge\star_5\D\theta_1 + \D\theta_2\wedge\star_5\D\theta_2) + \frac{1}{2}e^{4\varphi}(\theta_1 \D\theta_2 - \theta_2 \D\theta_1) \wedge\star_5 (\theta_1 \D\theta_2 - \theta_2 \D\theta_1)\\
        &\left.-\frac{1}{2}(3a_{IJ}+b_{IJ})F^I\wedge\star_5 F^J - (z A^1-A^2)\wedge\left(F^1\wedge F^1 - F^2 \wedge F^2\right) \right],
    \end{split}
\end{equation}
with
\begin{equation}
    b_{IJ} = \Sigma^4 \begin{pmatrix}
z^2 & -z \\
-z & 1 
\end{pmatrix}, \qquad a_{I J} = \frac{2}{3}\Sigma^{-2}\begin{pmatrix}
        &\cosh{2\alpha} &-\sinh{2\alpha}\\
        &-\sinh{2\alpha} &\cosh{2\alpha}
    \end{pmatrix},
\end{equation}
and
\begin{equation}\label{eq:fullpotential}
    \begin{split}
        V_5 = &\frac{1}{R^2}\biggl\{ \frac{e^{4\varphi}}{4\Sigma^4} - \frac{2e^{2\varphi}\cosh\alpha}{\Sigma} + \Sigma^2\Bigl[ -2+e^{2\varphi}(2\sinh^2\alpha (\theta_1^2 + \theta_2^2)+1) \\
        & +\frac{1}{8}e^{4\varphi}\bigl[ (\cosh(2\alpha)(2\theta_1^2 + 2\theta_2^2 +1)^2  -z (-z \cosh(2\alpha) + 2 \sinh(2\alpha) (2\theta_1^2 + \theta_2^2 +1)\bigl]  \Bigl] \biggl\}.
    \end{split}
\end{equation}
Setting $\theta_{1,2}=0$ together with a final redefinition of the gauge fields
\begin{equation}
    \begin{split}
        &A^1 = -\frac{e^{-2g_0/3}}{2^{7/3}\sqrt{3}}\left(2 \left(e^{2\lambda_A^{(0)}} - e^{2\lambda_B^{(0)}}\right)A^R - 3 e^{-f_0} A^F \right),\\
        &A^2 = -\frac{e^{-2g_0/3}}{2^{5/3}\sqrt{3}} \left(e^{2\lambda_A^{(0)}} + e^{2\lambda_B^{(0)}}\right) A^R,
    \end{split}
\end{equation}
brings us to the action presented in Section~\ref{sec:gravitytheory}.

%%%%%%%%%%%%%%%%%%%%%%%%%%%%%%
\section{Relation to the STU model for \texorpdfstring{$z=\pm1$}{}} \label{app:STU}
%%%%%%%%%%%%%%%%%%%%%%%%%%%%%%
In Section~\ref{sec:z=1bhsolutions} we studied the theory \eqref{eq:5dactionnotheta1} in the case of $z=\pm1$, where the hyperscalar is zero and the other two scalars are related via \eqref{eq:z=1scalars}. 
Here we show precisely how this theory is related to the STU model, whose action is
\begin{equation}\label{eq:STUaction}
    I_5 = \frac{1}{16\pi G_5}\int R_5 \star_51 - Q_{IJ}\mathcal{F}^I\wedge\star_5 \mathcal{F}^J - Q_{IJ}dX^I\wedge\star_5 dX^J - \frac{1}{6}C_{IJK}\mathcal{F}^I\wedge \mathcal{F}^J\wedge \mathcal{A}^K + 2m^2V_5\star_51,
\end{equation}
where $I,J,K=1,2,3$. The constants $C_{IJK}$ are set to 1 if $(IJK)$ is a permutation of $123$ and 0 otherwise. The three scalars $X^1,X^2,X^3$ are subject to a constraint
\begin{equation}\label{eq:STUscalarscondition}
    \frac{1}{6}C_{IJK}X^I X^J X^K = 1, \qquad \Rightarrow \qquad X^1 X^2 X^3 = 1.
\end{equation}
We define
\begin{equation}
    X_I \equiv \frac{1}{6}C_{IJK}X^J X^K, \qquad \Rightarrow \qquad X_I X^I = 1.
\end{equation}
The matrix $Q_{IJ}$ used for raising and lowering indices is defined as
\begin{equation}
    Q_{IJ} \equiv \frac{9}{2}X_I X_J - \frac{1}{2}C_{IJK}X^K = \frac{9}{2}\text{diag}((X_1)^2,(X_2)^2,(X_3)^2).
\end{equation}
Finally, the potential for the scalar(s) is given by
\begin{equation}
    V_5 = 27C^{IJK}\bar{X}_I \bar{X}_I X_K,
\end{equation}
where the constants $\bar{X}_I$ are the vacuum values of the scalars. It is easy to see that the action \eqref{eq:5dactionnotheta1}, upon setting the scalars to zero, is equivalent to \eqref{eq:STUaction} if we set
\begin{equation}
    \begin{split}
        &X^1 = 2^{2/3}\Sigma^4, \qquad \qquad \quad X^2 = X^3 = \frac{1}{2^{1/3}\Sigma^2},\\
        &\mathcal{A}^1 = \frac{-2 A^R-3 A^F}{2 \sqrt{3}}, \qquad \mathcal{A}^2 = \mathcal{A}^3 = \frac{-4 A^R+3 A^F}{4 \sqrt{3}},
    \end{split}
\end{equation}
for $z=1$ and
\begin{equation}
    \begin{split}
        &X^1 = 2^{2/3}\Sigma^4, \qquad \qquad \quad X^2 = X^3 = \frac{1}{2^{1/3}\Sigma^2},\\
        &\mathcal{A}^1 = \frac{-4 A^R-3 A^F}{2 \sqrt{3}}, \qquad \mathcal{A}^2 = \mathcal{A}^3 =\frac{-2 A^R+3 A^F}{4 \sqrt{3}},
    \end{split}
\end{equation}
for $z=-1$.

\bibliographystyle{JHEP}
\bibliography{HyperDraft}

\end{document}